\documentclass[apj,iop]{emulateapj}

\usepackage{graphics}
\usepackage{natbib}
\usepackage{color}

\newcommand{\snia}{SN~Ia}
\newcommand{\sneia}{SNe~Ia}
\newcommand{\snf}{SNfactory}
\newcommand{\ha}{H$\alpha$}
\newcommand{\hb}{H$\beta$}
\newcommand{\galex}{\emph{GALEX}}
\newcommand{\zpeg}{{\tt ZPEG}}

\newcommand{\nifs}{$^{56}$Ni}

\begin{document}

\title{Host Galaxies of Type Ia Supernovae from the Nearby Supernova
Factory}

\author
{
    M.~Childress,\altaffilmark{1,2,3,4}
    G.~Aldering,\altaffilmark{1}
    P.~Antilogus,\altaffilmark{5}
    C.~Aragon,\altaffilmark{1}
    S.~Bailey,\altaffilmark{1}
    C.~Baltay,\altaffilmark{6}
    S.~Bongard,\altaffilmark{5}
    C.~Buton,\altaffilmark{7}
    A.~Canto,\altaffilmark{5}
    F.~Cellier-Holzem,\altaffilmark{5}
    N.~Chotard,\altaffilmark{8}
    Y.~Copin,\altaffilmark{8}
    H.~K. Fakhouri,\altaffilmark{1,2}
    E.~Gangler,\altaffilmark{8}
    J.~Guy,\altaffilmark{5} 
    E.~Y. Hsiao,\altaffilmark{1}
    M.~Kerschhaggl,\altaffilmark{7}
    A.~G.~Kim,\altaffilmark{1}
    M.~Kowalski,\altaffilmark{7}
    S.~Loken,\altaffilmark{1,*}
    P.~Nugent,\altaffilmark{9}
    K.~Paech,\altaffilmark{7}
    R.~Pain,\altaffilmark{5}
    E.~Pecontal,\altaffilmark{10}
    R.~Pereira,\altaffilmark{8}
    S.~Perlmutter,\altaffilmark{1,2}
    D.~Rabinowitz,\altaffilmark{6}
    M.~Rigault,\altaffilmark{8}  
    K.~Runge,\altaffilmark{1}
    R.~Scalzo,\altaffilmark{3}
    G.~Smadja,\altaffilmark{8}
    C.~Tao,\altaffilmark{11,12}
    R.~C. Thomas,\altaffilmark{9}
    B.~A.~Weaver,\altaffilmark{13}
    C.~Wu\altaffilmark{5,14}
}

\altaffiltext{1}
{
    Physics Division, Lawrence Berkeley National Laboratory, 
    1 Cyclotron Road, Berkeley, CA, 94720.
}
\altaffiltext{2}
{
    Department of Physics, University of California Berkeley,
    366 LeConte Hall MC 7300, Berkeley, CA, 94720-7300.
}
\altaffiltext{3}
{
    Research School of Astronomy and Astrophysics,
    Australian National University,
    Canberra, ACT 2611, Australia.
}
\altaffiltext{4}
{
    ARC Centre of Excellence for All-sky Astrophysics (CAASTRO).
}
\altaffiltext{5}
{
    Laboratoire de Physique Nucl\'eaire et des Hautes \'Energies,
    Universit\'e Pierre et Marie Curie Paris 6, Universit\'e Paris Diderot Paris 7, CNRS-IN2P3, 
    4 place Jussieu, 75252 Paris Cedex 05, France.
}
\altaffiltext{6}
{
    Department of Physics, Yale University, 
    New Haven, CT, 06250-8121.
}
\altaffiltext{7}
{
    Physikalisches Institut, Universit\"at Bonn,
    Nu\ss allee 12, 53115 Bonn, Germany.
}
\altaffiltext{8}
{
    Universit\'e de Lyon, F-69622, Lyon, France ; Universit\'e de Lyon 1, Villeurbanne ; 
    CNRS/IN2P3, Institut de Physique Nucl\'eaire de Lyon.
}
\altaffiltext{9}
{
    Computational Cosmology Center, Computational Research Division, Lawrence Berkeley National Laboratory, 
    1 Cyclotron Road MS 50B-4206, Berkeley, CA, 94720.
}
\altaffiltext{10}
{
    Centre de Recherche Astronomique de Lyon, Universit\'e Lyon 1,
    9 Avenue Charles Andr\'e, 69561 Saint Genis Laval Cedex, France.
}
\altaffiltext{11}
{
    Centre de Physique des Particules de Marseille , Aix-Marseille Universit\'e , CNRS/IN2P3, 163, avenue de Luminy - Case 902 - 13288 Marseille Cedex 09, France.
}
\altaffiltext{12}
{
    Tsinghua Center for Astrophysics, Tsinghua University, Beijing 100084, China.
}
\altaffiltext{13}
{
    Center for Cosmology and Particle Physics,
    New York University,
    4 Washington Place, New York, NY 10003, USA.
}
\altaffiltext{14}
{
    National Astronomical Observatories, Chinese Academy of Sciences, Beijing 100012, China.
}
\altaffiltext{*}
{
    Deceased.
}

\begin{abstract}
We present photometric and spectroscopic observations of galaxies hosting Type Ia supernovae (\sneia) observed by the Nearby Supernova Factory (\snf). 
Combining \galex\ UV data with optical and near infrared photometry, we employ stellar population synthesis techniques to measure \snia\ host galaxy stellar masses, star-formation rates (SFRs), and reddening due to dust. We reinforce the key role of \galex\ UV data in deriving accurate estimates of galaxy SFRs and dust extinction.
Optical spectra of \snia\ host galaxies are fitted simultaneously for their stellar continua and emission lines fluxes, from which we derive high precision redshifts, gas-phase metallicities, and \ha-based SFRs. 
With these data we show that \snia\ host galaxies present tight agreement with the fiducial galaxy mass--metallicity relation from SDSS for stellar masses $log(M_*/M_\odot)>8.5$ where the relation is well-defined. The star-formation activity of \snia\ host galaxies is consistent with a sample of comparable SDSS field galaxies, though this comparison is limited by systematic uncertainties in SFR measurements.
Our analysis indicates that \snia\ host galaxies are, on average, typical representatives of normal field galaxies.
\end{abstract}

\keywords{supernovae: general}

\section{Introduction}
Type Ia supernovae (\sneia) are some of the most dramatic explosive events in the Universe, yet the exact nature of their stellar progenitors remains a mystery. \sneia\ are very bright, and exhibit low intrinsic luminosity dispersion \citep[$\sim0.35$~mag;][]{branchmiller93}, making them suitable for measuring extragalactic distances. The discovery of an empirical relationship between the luminosity and light curve decline rate of \sneia\ \citep{pskovskii77, phillips93, phillips99}, aided in large part by the discovery of extreme events such as SN~1991T \citep{phillips91T, filippenko91T} and SN~1991bg \citep{filippenko91bg, leibundgut91bg}, resulted in a significant improvement in the standardized luminosity of \sneia\ and enabled their successful use as cosmological probes. Discovery of a second empirical relationship between the observed color and luminosity of \sneia\ \citep{hamuy96, riess96, tripp98} further decreased the observed dispersion to the current level of about $\sim0.15$~mag. This uniformity of \snia\ luminosities has enabled their successful use as standardizable candles to constrain the energy content of the Universe \citep{riess98, perlmutter99, tonry03, knop03, mwv07, riess07, kowalski08, hicken09b, kessler09, union2, sullivan11a, suzuki12}. 

Despite their utility as cosmological distance indicators, the stellar progenitors of \sneia\ remain as yet undetermined. With future \snia\ surveys planned to find \sneia\ at high redshifts \citep[e.g. WFIRST --][]{wfirst}, concerns remain that the younger stellar ages and lower metallicities of high redshift environments could bias cosmological measurements if the corrected brightnesses of \sneia\ vary with these parameters. A promising source for clues to the origin of \snia\ brightness diversity, and its possible dependence on progenitor properties, is the study of \snia\ environments.

Early studies of \snia\ host galaxies found qualitative evidence for a correlation between the observed peak magnitude, light curve decline rate, and expansion velocity of an \snia\ with the morphological type of its host galaxy \citep{filippenko89, branchvdb93, hamuy96}, such that brighter slower declining \sneia\ preferentially occur in later type (spiral and irregular) galaxies, while fainter slower declining \sneia\ preferentially occur in earlier type (E/S0) galaxies. Recent studies have confirmed these trends and found analogous correlations with respect to host galaxy mass or metallicity \citep{hamuy00, gallagher05, gallagher08, neill09, howell09}. These trends provide compelling evidence that some property of the progenitor (most likely age or metallicity) that correlates with host galaxy properties may be driving the observed diversity of \snia\ light curve decline rates.

Recently, troubling evidence for progenitor-driven trends of \emph{corrected} \snia\ luminosities arose from studies of \snia\ Hubble residuals and the properties of their host galaxies \citep{kelly10, sullivan10, lampeitl10, gupta11, dandrea11, johansson12, hayden12}. Such a correlation can easily bias the measurement of cosmological parameters, particularly the dark energy equation of state parameter and its evolution with redshift. This subject will be the focus of the second paper in this series, but here it motivates a critical question: if a particular property of host galaxies should be used to correct \snia\ luminosities, what is the best technique for measuring both the value \emph{and uncertainty} of that property?

The most accessible physical properties to measure for galaxies are those which can be inferred from multi-band photometry, particularly stellar mass and star-formation rate (SFR). To derive those galaxy properties from photometry requires invocation of the art of stellar population synthesis (SPS). This is a rich field of study \citep[see][for a recent review]{conroy13}, to which we will later devote an entire section, and is being used with increasing frequency in the study of \snia\ host galaxies. SPS involves the comparison of observed galaxy photometry (across multiple filters) to fluxes predicted for a model spectral energy distribution (SED) generated for a chosen galaxy star-formation history (SFH). While much effort in the field of SPS is devoted to deriving model SEDs that best match observed galaxy photometry, we are here more interested in how much variation of the SFH is allowed for a galaxy given its observational measurement errors and a known level of uncertainties in the models. We will also inspect the effects of broader wavelength coverage in photometry, and will echo the findings of previous authors \citep{papovich01, gdpm02, neill09, gupta11} who showed that UV data is critical for accurately assessing a galaxy's recent star-formation activity.

While the photometrically derived properties of stellar mass and SFR are the most easily accessible galaxy properties, metallicity is a property of critical interest for \sneia. Galaxy metallicities are most directly measured from spectroscopy, but as this is observationally expensive a more desirable technique would be to infer metallicity from photometrically derived properties. This can be achieved by invoking galaxy scaling relations such as the galaxy mass-metallicity relation \citep[e.g.][]{trem04} or the mass-metallicity-SFR relation \citep{mannucci10a, laralopez10}. However, a critical check on this technique should be a confirmation that \snia\ host galaxies exhibit metallicities and SFRs consistent with expectations given their stellar masses, which we perform in this work.

This paper is the first in a series of papers which will study the host galaxies of \sneia\ discovered or followed by the Nearby Supernova Factory \citep[SNfactory,][]{ald02}. Here we present the observational data for the \snf\ host galaxy sample which will form the basis of the scientific analyses in this and future papers. In Section~\ref{sec:host_data} we present the photometric and spectroscopic observational data for our \snia\ host galaxy sample. We derive estimates of stellar mass, star-formation rate (SFR), and dust extinction for each host galaxy using a stellar population synthesis method described in detail in Section~\ref{sec:host_phot_sps}. We then exploit the derived host galaxy physical parameters (mass, metallicity, and SFR) to examine the question of whether \snia\ host galaxies are typical representatives of the normal galaxy population in the local universe, considering \snia\ host metallicities in Section~\ref{sec:snia_host_MZ} and host SFRs in Section~\ref{sec:snia_host_sfr}. We discuss the implications of our analyses in Section~\ref{sec:summary} and summarize our conclusions in Section~\ref{sec:conclusions}.

Throughout this paper we employ a standard $\Lambda$CDM cosmology with $\Omega_\Lambda = 0.7$, $\Omega_M = 0.3$, $H_0=70$~km~s$^{-1}$~Mpc$^{-1}$. Stellar masses and star-formation rates in this paper are computed using a \citet{chab03} initial mass function (IMF), and our prescriptions for converting literature values to this choice of IMF are described where appropriate.

\section{Host Galaxy Data}
\label{sec:host_data}
In this Section we present the observational data for the full sample of \snf\ \snia\ host galaxies. We first briefly describe the \snia\ sample in Section~\ref{sec:sn_sample}. In Section~\ref{sec:snf_host_phot} we describe the galaxy photometric data that was collected from numerous public sources, as well as the reduction of targeted observations using the \snf\ SNIFS instrument. Derivation of galaxy stellar masses and specific star formation rates utilizes a complex custom stellar population synthesis technique whose description is deferred to Section~\ref{sec:host_phot_sps}. Longslit spectroscopic data from numerous sources are described in Section~\ref{sec:snf_host_spec}, along with the derivation of gas-phase metallicities and \ha-based star formation rates.

\subsection{Supernova Sample}
\label{sec:sn_sample}
The \sneia\ whose hosts are analyzed here were observed as part of the ongoing science operations for the Nearby Supernova Factory \citep[\snf\ --][]{ald02}. The \snf\ was designed to observe several hundred \sneia\ in the nearby smooth Hubble flow ($0.03 < z < 0.08$) with the goals of achieving a deeper physical understanding of \sneia, building better \snia\ templates for cosmological applications, and anchoring the low-redshift Hubble Diagram. We briefly summarize the details of these operations.

From 2005 to 2008, the \snf\ conducted a wide-field search of the northern and equatorial sky using the QUEST-II CCD camera \citep{baltay07} on the Samuel Oschin 1.2m Schmidt telescope on Mount Palomar, California, partly in collaboration with the JPL Near-Earth Asteroid Tracking (NEAT) component of the Palomar-QUEST consortium. Typical search images consisted of 60~s exposure with an RG610 filter, with each field revisited multiple times in order to detect asteroids (which we reject). The \snf\ search covered an average unique area of 600 deg$^2$ per night and covered nearly half the sky ($\approx$20,000 deg$^2$) each year. In 28 months of searching, the \snf\ discovered over 1000 supernovae of all types, and spectroscopically confirmed over 600 of those. 

Spectroscopic typing of search candidates and followup observations of \sneia\ were obtained with the SuperNova Integral Field Spectrograph \citep[SNIFS --][]{ald02,lantz04}, mounted continuously on the University of Hawaii 2.2-m telescope on Mauna Kea. In addition to its spectroscopic capabilities, SNIFS has an imaging channel which was used to obtain broadband photometry of some \snia\ host galaxies in our sample (see Section~\ref{sec:snifs_phot}).

A total of 400 supernovae discovered with the \snf\ search were spectroscopically confirmed to be \sneia, and those \sneia\ discovered before B-band maximum light (as estimated by spectroscopic typing) were followed up extensively with SNIFS. In addition to those \sneia\ discovered by \snf, some \sneia\ discovered by other searches were followed with SNIFS, especially during times when search operations were impacted by weather or fires on Mount Palomar. This work analyzes all \sneia\ discovered or followed by \snf, a total of 469 \sneia\ observed from 2004-2010. Of these \sneia, a total of 216 were extensively followed with SNIFS, with 185 of those being discovered by the \snf\ search.

In Table~\ref{tab:host_info} we present the location of the \snia\ host galaxy as determined from the $g$-band host image (see Section~\ref{sec:final_host_phot}), the redshift measured from the host spectrum (see Section~\ref{sec:host_redshifts}) for all \sneia\ in our sample, and the discovery source for \sneia\ not discovered by \snf.  Most of the SNe in our sample which were not discovered by \snf\ were discovered by amateur astronomers, the Palomar Transient Factory \citep[PTF][]{ptf09}, the Lick Observatory Supernova Search \citep[LOSS][]{LOSS}, or the ROTSE supernova search \citep{rotse, quimbythesis, yuanthesis}.

\begin{table*}
\begin{center}
\scriptsize
\caption{\snf\ \snia\ Host Galaxy Properties}
\label{tab:host_info}
\begin{tabular}{lllll}
\hline
SN Name  & Host Coords  & Redshift (Helio.) & Reference & Discoverer \\
\hline
SNF20050519-000 & 21:35:52.58 -26:27:06.1 & $0.03040 \pm 0.00500$ &    & SNF\\
SNF20050621-001 & 20:45:17.18 -03:49:37.5 & $0.12072 \pm 0.00011$ &    & SNF\\
SNF20050624-000 & 17:34:46.48 -02:21:26.2 & $0.06719 \pm 0.00009$ &    & SNF\\
SNF20050704-008 & 22:26:29.49 -16:41:38.4 & $0.11116 \pm 0.00005$ &    & SNF\\
SNF20050727-005 & 21:09:08.78 -15:38:45.9 & $0.08610 \pm 0.00007$ &    & SNF\\
SNF20050728-000 & 00:57:29.59 -22:35:15.3 & $0.04293 \pm 0.00009$ &    & SNF\\
SNF20050728-001 & 00:11:17.78 -28:54:45.3 & $0.05651 \pm 0.00010$ &  1 & SNF\\
SNF20050728-006 & 21:58:04.87 -14:00:01.8 & $0.05921 \pm 0.00007$ &    & SNF\\
SNF20050728-012 & 23:56:29.32 -22:49:32.1 & $0.07477 \pm 0.00250$ &    & SNF\\
SNF20050729-008 & 22:06:56.95 -09:29:30.4 & $0.08206 \pm 0.00005$ &    & SNF\\
\hline
\end{tabular}
\small
\\
{\bf References}: (1) \citet{zSNF20050728001}
\normalsize
\end{center}
\end{table*}

A total of 40 SNe in our sample have been presented in other \snia\ samples. 27 of our sample were also part of the CfA \snia\ sample \citep{hicken09b, hicken12}, 24 were observed by the Carnegie Supernova Project \citep[CSP --][]{contreras10, stritzinger11}, 21 were observed by LOSS \citep{ganesh10}, and 1 (SN~2005hc) was part of the \citet{gupta11} sample. Host galaxy masses were previously derived for 1 (SN~2006X) \snia\ host by \citet{neill09}, but we did not derive a host mass for this galaxy due to shredding in photometric source extraction. 14 of our \snia\ hosts have masses derived by \citet{kelly10}, and we found our values (see Section~\ref{sec:host_phot_sps}) had good agreement with theirs, having a mean offset of 0.04~dex in stellar mass.

\subsection{Host Galaxy Photometric Data}
\label{sec:snf_host_phot}
Photometric data for \snf\ \snia\ host galaxies was gathered from public sources as well as our own targeted observations. Optical photometry was collected from the Sloan Digital Sky Survey \citep[SDSS,][]{york00} Eighth Data Release \citep[DR8,][]{aihara11}. NIR images from the Two Micron All Sky Survey \citep[2MASS,][]{twomass} were obtained at the NASA/IPAC Infrared Science Archive (IRSA\footnote{http://irsa.ipac.caltech.edu}), and for select hosts NIR data were available from the UKIRT Infrared Deep Sky Survey \citep[UKIDSS,][]{ukidss}. UV data taken with the Galaxy Evolution Explorer \citep[\galex,][]{galex} satellite were obtained from the \galex\ online data archive at MAST\footnote{http://galex.stsci.edu}.

The public photometric coverage of our hosts was very good.  344 of the 469 hosts (73\%) fell within the SDSS photometric footprint, all 469 hosts had 2MASS imaging data, 58 (12\%) hosts had UKIDSS data, and 408 (87\%) have \galex\ AIS (All-Sky Imaging Survey - a shallow imaging survey) images. Additionally, 102 (22\%) of our hosts have deeper \galex\ imaging, mostly from the MIS (Medium Imaging Survey).

The typical photometric depth for these surveys (for this work, this limit is effectively where the flux errors reach 5-10\% for point sources) are 20th magnitude for SDSS, 17th magnitude for 2MASS, 20th magnitude for UKIDSS, 19th magnitude for \galex\ AIS, and 21st magnitude for \galex\ MIS. After these cuts, the total number of hosts with data from each survey is 323 (69\%) from SDSS, 394 (84\%) from \galex, 272 (58\%) from 2MASS, and 49 (10\%) from UKIDSS.

For those hosts without optical photometry from SDSS, we used SNIFS in imaging mode to obtain optical images of the host after the SN had faded (at least one year later).  SNIFS was also used to obtain deeper photometry for those faint hosts whose SDSS images were not deep enough (typically for $m_g > 19.0$).  A total of 283 hosts (60\% of the total sample of 469 hosts) have SNIFS photometry. For 10 hosts, $g$-band photometry was obtained with Keck LRIS prior to spectroscopic observations (see Section~\ref{sec:snf_host_spec}) of the hosts, and was later zero-pointed to either SDSS or SNIFS photometry.

\begin{figure*}
\begin{center}
\includegraphics[width=0.90\textwidth]{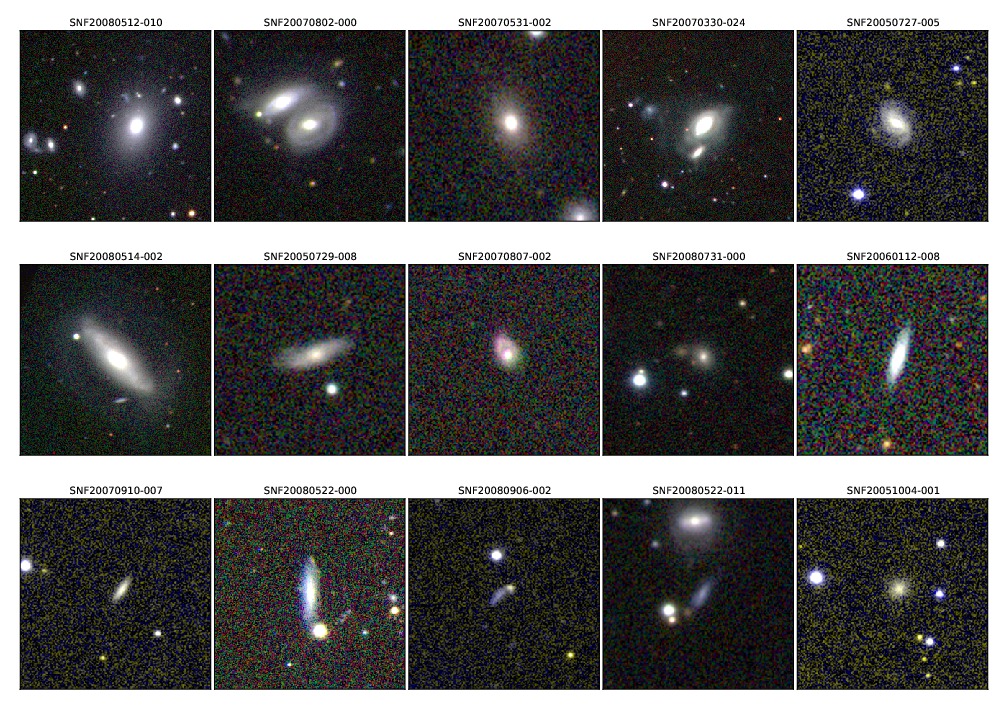}
\end{center}
\caption[\snf\ Host Mosiac]{A subsample of the \snf\ host galaxies, presented in $gri$ color composites constructed using {\tt STIFF} \citep{stiff}. Galaxies are order by stellar mass from highest (upper left) to lowest (lower right).}
\label{fig:snf_host_mosaic}
\end{figure*}

Below we describe our reduction of the SNIFS photometry and our method of combining multi-band imaging data to obtain accurate common aperture photometry. Our means of deriving galaxy stellar masses and star formation rates from photometry is described in Section~\ref{sec:host_phot_sps}. A mosaic image of several \snf\ host galaxies is shown in Figure~\ref{fig:snf_host_mosaic}, utilizing both SDSS and SNIFS optical data and spanning a large range (4 orders of magnitude) of stellar masses.

\subsubsection{SNIFS Photometry}
\label{sec:snifs_phot}
\begin{figure}
\begin{center}
\includegraphics[width=0.45\textwidth]{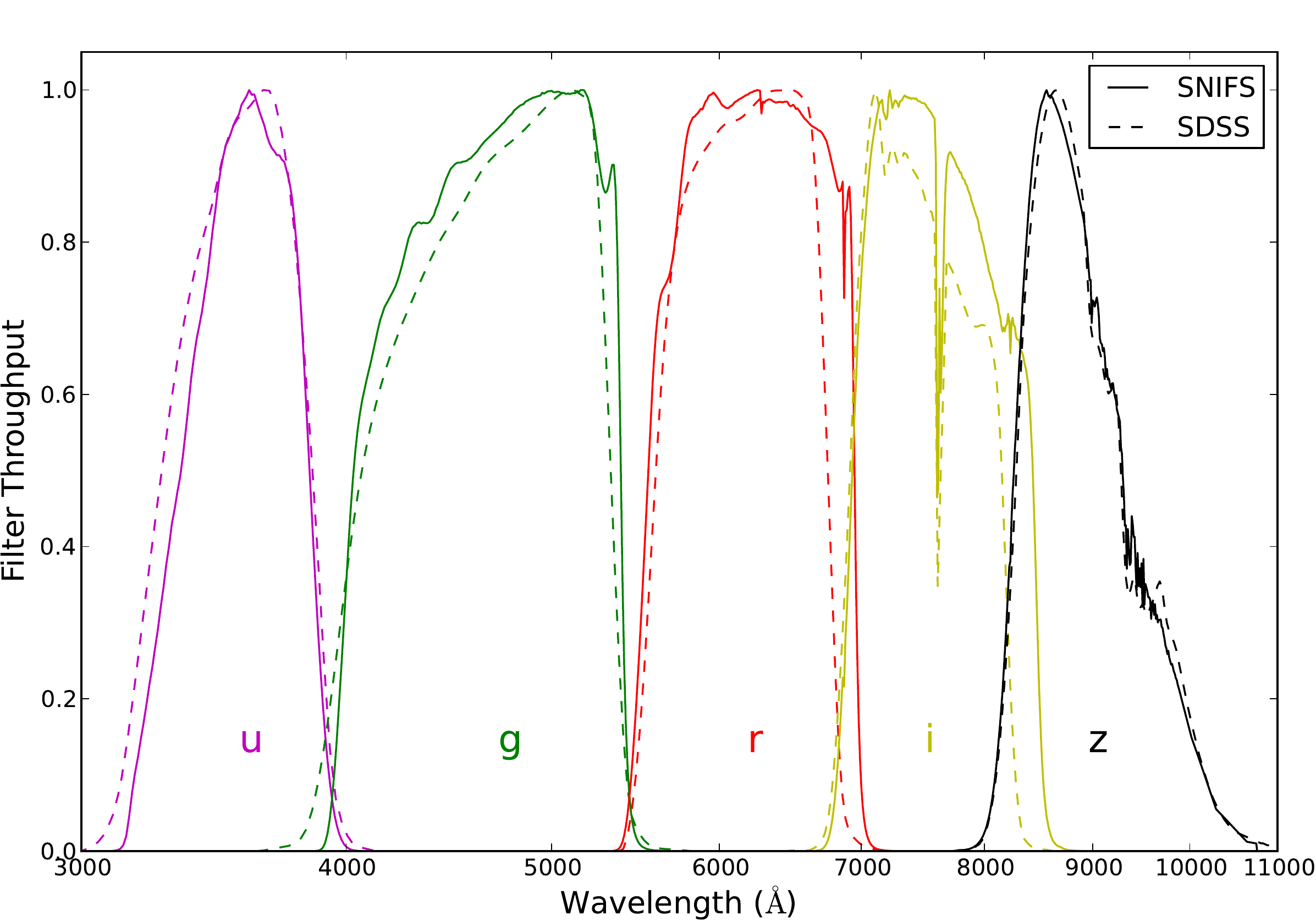}
\end{center}
\caption[SNIFS filter throughputs vs. SDSS.]{SNIFS filter throughputs, compared to those of SDSS.}
\label{fig:snifs_filts}
\end{figure}

For those hosts without publicly available optical photometry from SDSS, or those faint hosts for which the photometric depth of SDSS was insufficient, we obtained optical photometry using SNIFS in imaging mode.  The photometric imager (``P-channel'') on SNIFS consists of two 2k$\times$4k E2V CCDs, with one ``guider'' chip undergoing fast continuous windowed readout to perform guiding during observations, and the other ``science'' chip dedicated to photometry. In normal SNIFS \snia\ observation mode, the P-channel uses our custom ``multi-filter,'' which observes field stars on the science chip segmented into multiple filters spanning a broad wavelength range, in order to estimate atmospheric extinction. The SNIFS P-channel is also equipped with a variety of broadband filters covering the full science chip, including the standard Gunn $ugriz$ filters employed by SDSS.  Because the SNIFS P-channel CCDs are different from those on the SDSS imager, the effective SNIFS filter throughputs vary slightly from those of SDSS.  We show in Figure~\ref{fig:snifs_filts} the fiducial SNIFS filter throughputs derived from the throughput of all the optical components involved and our Mauna Kea extinction curve \citep{buton13} compared to the SDSS filter throughputs.

SNIFS images were processed in 
IRAF\footnote{IRAF is distributed by the National Optical Astronomy Observatory which is operated by the Association of Universities for Research in Astronomy, Inc., under cooperative agreement with the National Science Foundation.}
using standard techniques.  Overscan subtraction was performed for both amplifiers on the science chip, and data from each amp scaled by its gain. Images were trimmed to remove vignetting by the filter casing, then divided by normalized flatfield dome images to remove pixel variations in detector efficiency. For the reddest filters ($i$- and $z$-band), fringe patterns were removed by scaling a master fringe frame to the fringing measured in sky pixels for each science image. Master fringe frames were constructed from numerous long exposures, and identification of sky pixels and fringe scaling were performed using custom software. Cosmic rays were then removed using {\tt LA Cosmic} \citep{vandokkum01}. Astrometric solutions for all images were derived using {\tt WCSTools} \citep{mink06}, then refined using {\tt SCAMP} \citep{scamp} matching to 2MASS \citep{twomass} astrometry. Images from fields with multiple exposures were combined with {\tt SWARP} \citep{swarp} using median addition.  For the purpose of deriving photometric zeropoints and host galaxy photometry (see Section~\ref{sec:final_host_phot}), fluxes for all objects in each image were measured with {\tt SExtractor} \citep{sextractor} using the {\tt FLUX\_AUTO} parameter.

The observing priorities for the SNIFS host photometry program were to obtain $g$-band and $i$-band photometry of all our hosts. The optical $g-i$ color is a very good color for determining galaxy mass-to-light ratios \citep{gb09} and thus serves as a minimal filter set for obtaining accurate galaxy masses. Many observations were taken between the Seventh Data Release \citep[DR7,][]{abaz09} and Eighth Data Release \citep[DR8,][]{aihara11} of SDSS, which added a significant area to the SDSS imaging footprint. Thus we have a large number of fields in the SDSS footprint observed by SNIFS, especially in $g$ and $i$, and with many of those observed on photometric nights when photometric calibration solutions were derived. This enables both the study of SNIFS-SDSS color terms as well as an independent measurement of the accuracy of our photometric calibrations, and we describe these two studies below.

Photometric zeropoints for SNIFS imaging in the SDSS footprint were obtained by matching photometric measurements of field stars from each SNIFS science image to their values in the SDSS DR8 catalog. Formal zeropoints and their uncertainties were derived as the weighted mean (weighted by photometric error) of the zeropoints for individual field stars after the exclusion of severe outliers. In Table~\ref{tab:color_terms} we summarize the total number of SDSS fields visited with SNIFS in each band $N_{fields}$, the mean of the zeropoint uncertainties $\left<\sigma_{ZP}\right>$ for all SNIFS fields zeropointed with SDSS in that filter, and the total number of stars matched over all fields $N_{stars}$. Most of our zeropoints are derived from $\approx$ 30 field stars per image and have a precision of 0.01-0.02~mag.

As stated above, the filter throughputs from SNIFS differ slightly from those of SDSS, so we might expect small but nonzero color terms between the two filter sets. We can measure these from the same field stars used for zeropointing SNIFS photometry in the SDSS footprint. To do so, we compare the residual magnitude offsets (after application of the fitted zeropoint) of these field stars as a function of their color as measured by SDSS. We derive the weighted mean offsets in bins of color (typically 0.2~mag wide) and perform a minimization to derive the optimal color term and its uncertainty for each filter. These are summarized in Table~\ref{tab:color_terms}. As can be seen, the color terms are consistent with zero for all of the filters except $i$-band, which has a small but significant detection of a color term. This may be due to different strength of telluric absorption features at the SNIFS site (Mauna Kea) compared to the SDSS site (Apache Peak), or may be due to the slightly different red wavelength roll-off of the filter throughputs, as both of these effects may alter the shape of the filter throughput and drive a color term between SNIFS and SDSS.

\begin{table}[h]
\centering
\caption{SNIFS color terms}
\centering
\begin{tabular}{ l r r r r l }
\hline
Filter & $N_{fields}$ & $N_{stars}$ & $\left<\sigma_{ZP}\right>$  & Color Term & Color \\
\hline
$u$     &   9 &   192 & 0.0185 & $-0.0009 \pm 0.0269$ & $u-g$ \\
$g$     & 160 &  4914 & 0.0094 & $ 0.0004 \pm 0.0087$ & $g-r$ \\
$r$     &  12 &   790 & 0.0109 & $ 0.0014 \pm 0.0104$ & $g-r$ \\
$i$     & 157 & 12452 & 0.0143 & $-0.0222 \pm 0.0115$ & $r-i$ \\
$z$     &  12 &  1068 & 0.0294 & $ 0.0081 \pm 0.0561$ & $i-z$ \\
\hline        
$g$     &  -- &    -- &     -- & $-0.0010 \pm 0.0044$ & $g-i$ \\
$i$     &  -- &    -- &     -- & $ 0.0099 \pm 0.0056$ & $g-i$ \\
\hline
\end{tabular}
\label{tab:color_terms}
\end{table}

Photometric zeropoints for SNIFS fields outside the SDSS footprint were derived for each observing night in each filter using observations of standard stars spanning a large range of airmasses. Our standards were selected from the \citet{smith02} sample, placing our measurement on the standard $ugriz$ system employed by SDSS. For each night (in each filter) we fit for a global zeropoint and an atmospheric extinction term, and our extinction terms were consistent with those predicted by the fiducial Mauna Kea extinction curve \citep{buton13}. Typical dispersion of standard star magnitudes about the best fit calibration solution were about 0.02~mag in $gri$ and 0.03~mag in $u$ and $z$. New science images were assigned a zeropoint based on their airmass and exposure time as calculated with the fitted extinction solutions.

As stated above, a number of the fields for which we obtained new zeropoints were included in the subsequent SDSS data release, enabling us to derive external zeropoints to cross-check our calibration solutions. We compared the SNIFS-based zeropoints to those derived by matching to SDSS and found good agreement (mean zeropoint offsets less than about 0.005~mag) with a dispersion consistent with the dispersion seen in our calibration solutions (about 0.02-0.03~mag). Since the SDSS zeropoints are derived from a large number of field stars, they are ultimately more precise than those derived from SNIFS observations of a few standard stars. We thus use SDSS zeropoints in favor of SNIFS zeropoints where available.

\subsubsection{Final Host Galaxy Photometry}
\label{sec:final_host_phot}
With the final processed SNIFS imaging and public data from SDSS, 2MASS, UKIDSS, and \galex, we obtain magnitudes for our hosts in each band by performing common aperture photometry.  We use the $g$-band image to select the host galaxy and define the galaxy aperture. For each SN, the location and elliptical apertures of nearby galaxies were measured with {\tt SExtractor} \citep{sextractor} and the separation of each galaxy from the SN location was calculated and then scaled to the effective galaxy radius projected along the SN-galaxy vector. The SN host was then nominally identified as the galaxy the least number of effective radii from the SN location. A spectroscopic redshift of the prime galaxy candidate was required to match the SN redshift as determined using SNID \citep{snid} to within 1500 km s$^{-1}$. In nearly all instances the nearest host candidate was confirmed to be the host, though there were a few cases in which the nearest host candidate was a background high-redshift galaxy.

For some of the PTF targets from 2010, we will report preliminary host properties in Section~\ref{sec:host_phot_sps} based on the SN redshift and the nominal host galaxy. Some SNe appear to be hostless, and these interesting cases will be a subject of future study.

\begin{figure}
\begin{center}
\includegraphics[width=0.45\textwidth]{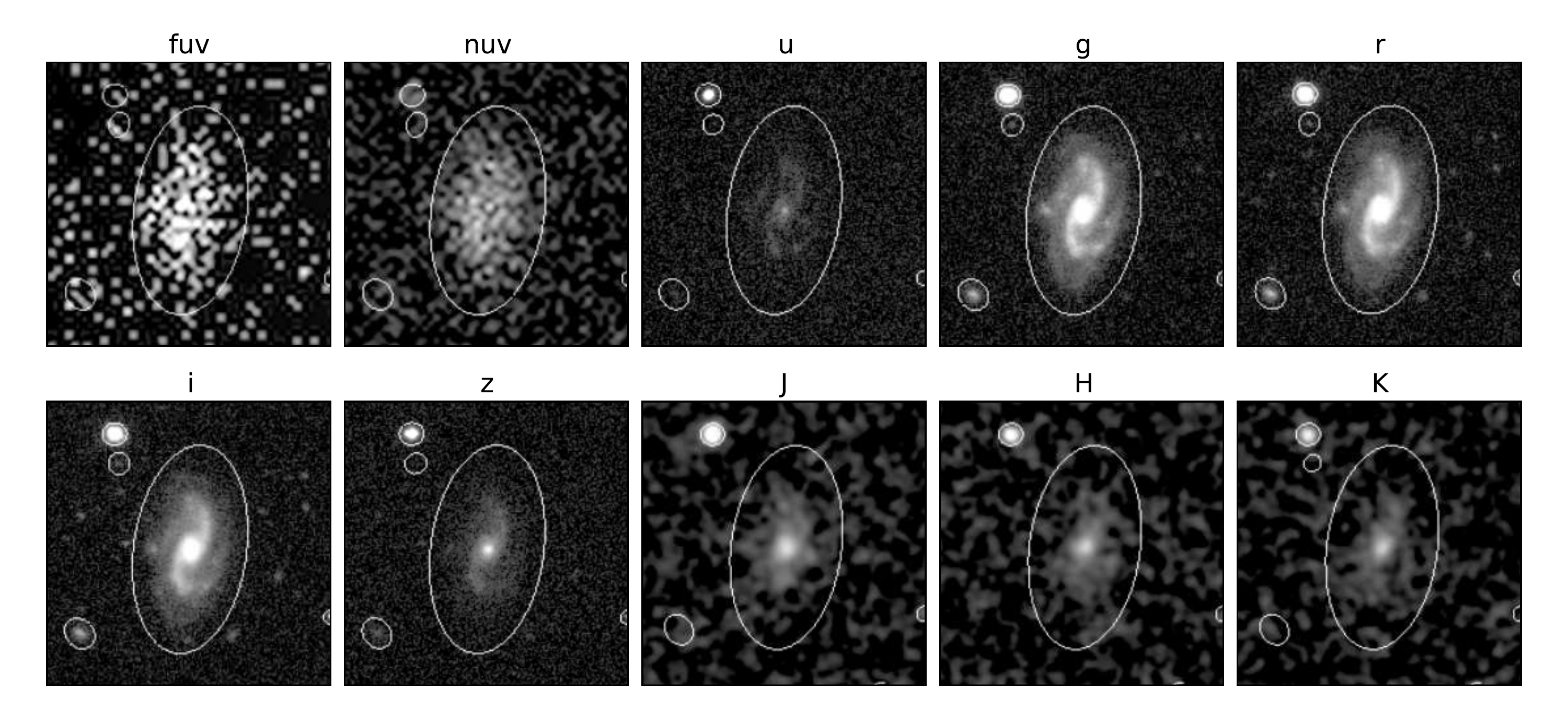}
\end{center}
\caption{Example of common aperture photometry for the host of SNF20060609-002, showing the resampled images in each band and the photometric aperture.}
\label{fig:common_aperture_example}
\end{figure}

After positive host identification, we measure the host flux in each band by first resampling the image from each filter to the resolution of the aperture image using {\tt SWARP} and then running {\tt SExtractor} \citep{sextractor} in dual image mode. We use the {\tt SExtractor} {\tt FLUX\_AUTO} output parameter, which measures the flux inside an elliptical Kron-like aperture. Sky background subtraction was performed using a large background mesh (1024 pixels, equivalent to about 4 arcmin). Star contamination of galaxy photometry was handled with {\tt SExtractor} object deblending, which assigns pixels to overlapping objects based on where the flux of the star drops below a certain contrast threshold, which we chose to be the default fraction of 0.005. An example of our common aperture photometry method is shown in Figure~\ref{fig:common_aperture_example}.

Galaxy magnitudes and their errors in each band were calculated using the zeropoints and noise characteristics for each image. We convert all magnitudes to the AB systems by applying Vega-AB offsets for 2MASS and UKIDSS magnitudes \citep{hewett06}, and the $ugriz$ offsets for SDSS as derived by \citet{kessler09}. \galex\ zeropoints are already on the AB system. Observed magnitudes were then corrected for foreground Milky Way reddening using the dust maps of \citet{sfd98} and the reddening law of \citet[][hereafter CCM]{ccm}.

For some images, such as \galex\ UV images of elliptical hosts with no star-formation, the galaxy flux was below the image noise threshold. For these images we calculate an effective $3\sigma$ magnitude upper limit determined by the flux uncertainty measured in the galaxy aperture after accounting for Milky Way dust reddening. These upper limits, especially those from \galex\ UV data, proved valuable in constraining the stellar population fits to photometry, as outlined in Section~\ref{sec:uv_limits}.

Finally, optical magnitudes were corrected for emission line fluxes using the fitted line profiles and stellar continuum measurements obtained from optical spectra (see Section~\ref{sec:snf_host_spec}) where available. These emission corrections were typically less than 0.01~mag in all bands except for $r$ or $i$ (depending on redshift) where the strongest emission line (\ha) was typically present. For the most strongly star-forming hosts in our sample, the emission corrections still were less than 0.05~mag.

To convert these galaxy photometric data into physical properties of the galaxies, one must invoke the art of stellar population synthesis (SPS). Because this is a rich and constantly evolving field, we devote the entirety of Section~\ref{sec:host_phot_sps} to describing our chosen method of deriving galaxy stellar masses and SFRs.

\subsection{Host Galaxy Spectroscopic Data}
\label{sec:snf_host_spec}
Galaxy spectroscopy is useful for gaining finer insight into the galaxy SED than can be gleaned from broadband photometry. In particular, absorption features in the stellar continuum of the galaxy SED can be compared to stellar evolution models to estimate stellar age and metallicity, while narrow emission lines from ionized HII regions surrounding young stars can yield both gas-phase metallicity of the galaxy interstellar medium (ISM) as well as the current rate of star formation. Additionally, reddening in the galaxy ISM and near the ionized HII regions can be estimated from galaxy spectra. In this Section we describe both the \snf\ host galaxy spectroscopic data set as well as the extraction of galaxy physical parameters from these data.

\subsubsection{\snf\ Host Spectroscopy Observations}
Longslit spectra for our \snia\ host galaxies were obtained during numerous observing runs at multiple telescopes from 2007-2011.  The instruments used were the Kast Double Spectrograph \citep{kast} on the Shane 3-m telescope at Lick Observatory, the Low Resolution Imaging Spectrometer \citep[LRIS --][]{oke95} on the Keck I 10-m telescope on Mauna Kea, the R-C Spectrograph 
on the Blanco 4-m telescope at Cerro Tololo Inter-American Observatory, the Goodman High Throughput Spectrograph \citep{clemens04} on the Southern Astrophysical Research (SOAR) 4-m telescope on Cerro Pachon, and GMOS-S \citep{davies97} on the Gemini-S 8-m telescope on Cerro Pachon. The instrument configurations, including wavelength coverage and effective resolution (FWHM), are presented in Table~\ref{tab:instruments}. Objects were generally aligned onto the slit using direct imaging techniques where available, or offset star alignment when direct imaging was not available. The slit position angle was generally chosen to correspond to the parallactic angle at the time of observation, though occasionally when the airmass was favorable (i.e. less than about 1.05) the slit was aligned along the galaxy-SN direction to enable future abundance gradient or rotation studies.

\begin{table*}[ht]
\centering
\caption{Instrument Configurations}
\centering
\begin{tabular}{ l c c c c c }
\hline
Instrument & Dichroic/ & Disperser & Slit     & Wavelength    & Effective \\
           & Filter    &           & (arcsec) & Coverage (\AA)& Resolution (\AA)\\
\hline
Kast blue   & d55   & 600/4310 & 2.0 & 3900-5550  &  3.1 \\
Kast red    & d55   & 300/7500 & 2.0 & 5450-10500 &  9.1 \\
            &       &          &     &            &      \\
LRIS blue   & D560  & 600/4000 & 1.0 & 3500-5600  &  3.9 \\
LRIS red    & D560  & 900/5500 & 1.0 & 5500-7650  &  4.2 \\
            &       &          &     &            &      \\
Goodman HTS & GG385 & 300 l/mm & 1.0 & 3850-7700  & 13.7 \\
            &       &          &     &            &      \\
R-C Spec    & GG385 & 300/7500 & 1.0 & 3850-7700  &  9.1 \\
            &       &          &     &            &      \\
GMOS-S      & GG455 & B600     & 1.5 & 5040-7920  &  6.8 \\
\hline
\end{tabular}
\label{tab:instruments}
\end{table*}

Longslit spectra were reduced in IRAF using standard techniques.  After overscan subtraction, we subtracted bias frames from two-dimensional longslit data, removed cosmic rays using {\tt LA Cosmic} \citep{vandokkum01}, and flatfielded to remove pixel variations in detector efficiency. Two-dimensional wavelength solutions were derived from arc lamp exposures taken either at the same pointing as the object spectrum (for Shane, Blanco, and SOAR data) or using nightly arc lamp exposures (for Keck and Gemini-S data), with a one-dimensional shift applied by measuring atomic (OI) night sky lines in object spectra.  Object spectra were reduced to one dimension using the IRAF function {\tt apall}, with the galaxy aperture chosen by visual inspection to encompass as much of the galaxy core as was available above the sky noise. Nightly flux calibrations were derived from standard stars observed at appropriate ranges of airmass, with an atmospheric extinction solution derived for each observing night. Telluric absorption features were then removed using the nightly standard star spectra, accounting for appropriate dependence on airmass. Observer motion with respect to the heliocentric frame was then corrected, and finally spectra were dereddened to correct for Milky Way extinction using the dust maps of \citet{sfd98} and the reddening law of \citet{ccm}.

Some hosts had spectra available from SDSS DR8 \citep{aihara11}. These spectra were downloaded and then converted to air wavelengths for consistency with reduction of our own observations. The total number of spectra from each source was 225 host spectra from Kast, 82 from SDSS, 29 from the R-C Spectrograph, 18 from LRIS, 13 from the Goodman HTS, and 7 from GMOS-S, for a total of 374 host spectra (81\% of the sample with host photometry). In Table~\ref{tab:host_specobs} we list the instrument, UTC date of observation, exposure time, slit PA, and mean airmass for each \snia\ host spectroscopic observation. For SDSS spectra, we list the plate-mjd-fiber combination of the spectroscopic observation in the observation date column.

\begin{table*}
\begin{center}
\caption{\snf\ \snia\ Host Spectroscopic Observations}
\label{tab:host_specobs}
\begin{tabular}{lccccc}
\hline
SN Name & Instrument & UTC Date & Exp. Time & Slit PA & Airmass \\
\hline
SNF20050621-001 & Kast & 2007-Oct-15.2 & 1800 & 111 & 1.499 \\
SNF20050624-000 & Kast & 2007-Jun-15.4 & 1800 & 82 & 1.326 \\
SNF20050704-008 & R-C Spec. & 2009-Aug-28.2 & 900 & 100 & 1.029 \\
SNF20050727-005 & R-C Spec. & 2009-Aug-30.1 & 1200 & 48 & 1.091 \\
SNF20050728-000 & Kast & 2007-Oct-14.3 & 1800 & 178 & 1.992 \\
SNF20050728-006 & Kast & 2010-Sep-11.3 & 1200 & 188 & 1.614 \\
SNF20050728-012 & R-C Spec. & 2009-Aug-29.2 & 900 & 189 & 1.010 \\
SNF20050729-008 & SDSS & 0718-52206-0264 & 9211 & \nodata & 1.443 \\
SNF20050730-003 & Kast & 2009-Oct-24.2 & 600 & 182 & 1.309 \\
SNF20050731-005 & Kast & 2009-Aug-21.3 & 3600 & 107 & 1.314 \\
\hline
\end{tabular}
\end{center}
\end{table*}

We note here that our visual selection of spectrum extraction apertures may result in a different distribution of spectroscopic covering fraction as compared to other spectroscopic surveys such as SDSS. For example, the typical extraction aperture used for our Kast observations was 10\arcsec\ with a 2\arcsec-wide slit, whereas the SDSS spectroscopic fibers covered a 3\arcsec\ diameter circular aperture. For the metallicity and SFR comparisons undertaken in Sections~\ref{sec:snia_host_MZ} and \ref{sec:snia_host_sfr}, we are interested in how our aperture selections might affect the metallicity and SFR values measured from emission line fluxes in our host galaxy spectra.

Differences between galaxy metallicity or line emission measured with different apertures have been studied by a number of authors, e.g., \citet{gomez03, kewley05, salim07, moustakas10, zahid13}. These differences depend on covering fraction, but in different ways for different measurements. In particular, \ha\ flux can vary strongly with position due to HII regions, thus affecting \ha-based SFR and equivalent width (EW) measurements, while metallicities will vary less strongly with position, e.g. primarily due to radial gradients. For instance \citet{gerssen12} find that SDSS near-core \ha\ emission has a poor correlation with global \ha\ for captured-light fractions below 40\%. \citet{kewley05} find that for slit spectroscopy a captured-light fraction greater than $\sim$20\% is adequate to achieve agreement between global and slit metallicities.

For a more quantitative estimate of the effect that galaxy metallicity gradients and aperture choice may have on our results, we turn to the recent study by \citet{sanchez12}. They found that metallicity gradients in spiral galaxies are very consistent across all spiral galaxy sub-types when scaled by the effective radius ($r_e$) of the galaxy, with mean and scatter of $-0.12\pm0.11$~dex$/r_e$. Our apertures cover typically one scale radius, so the light-weighted metallicity for an exponential disk would be only $0.03\pm0.03$~dex lower than the at-core metallicity. Since our metallicities are measured from emission lines rather than starlight, the light profile of the emission provides a more accurate result. \citet{koopmann06} found the \ha\ scale radius to be 1.14$\times$ the stellar scale radius on average. This implies that our emission-weighted metallicities would be only $0.04\pm0.04$~dex lower than the core metallicities. Even if integrated to infinite radius, our metallicities would be only $0.08\pm0.08$~dex lower than core for a emission-weighted extraction of an exponential disk. These potential offsets are well below the systematic uncertainties on the metallicity methods themselves and thus are unlikely to influence our results. This notion is supported by the fact that our hosts show such excellent agreement with the galaxy mass-metallicity relation (Section~\ref{sec:snia_host_MZ}).

\subsubsection{Redshifts and Emission Line Fluxes}
\label{sec:host_redshifts}
\begin{figure*}
\begin{center}
\includegraphics[width=0.90\textwidth]{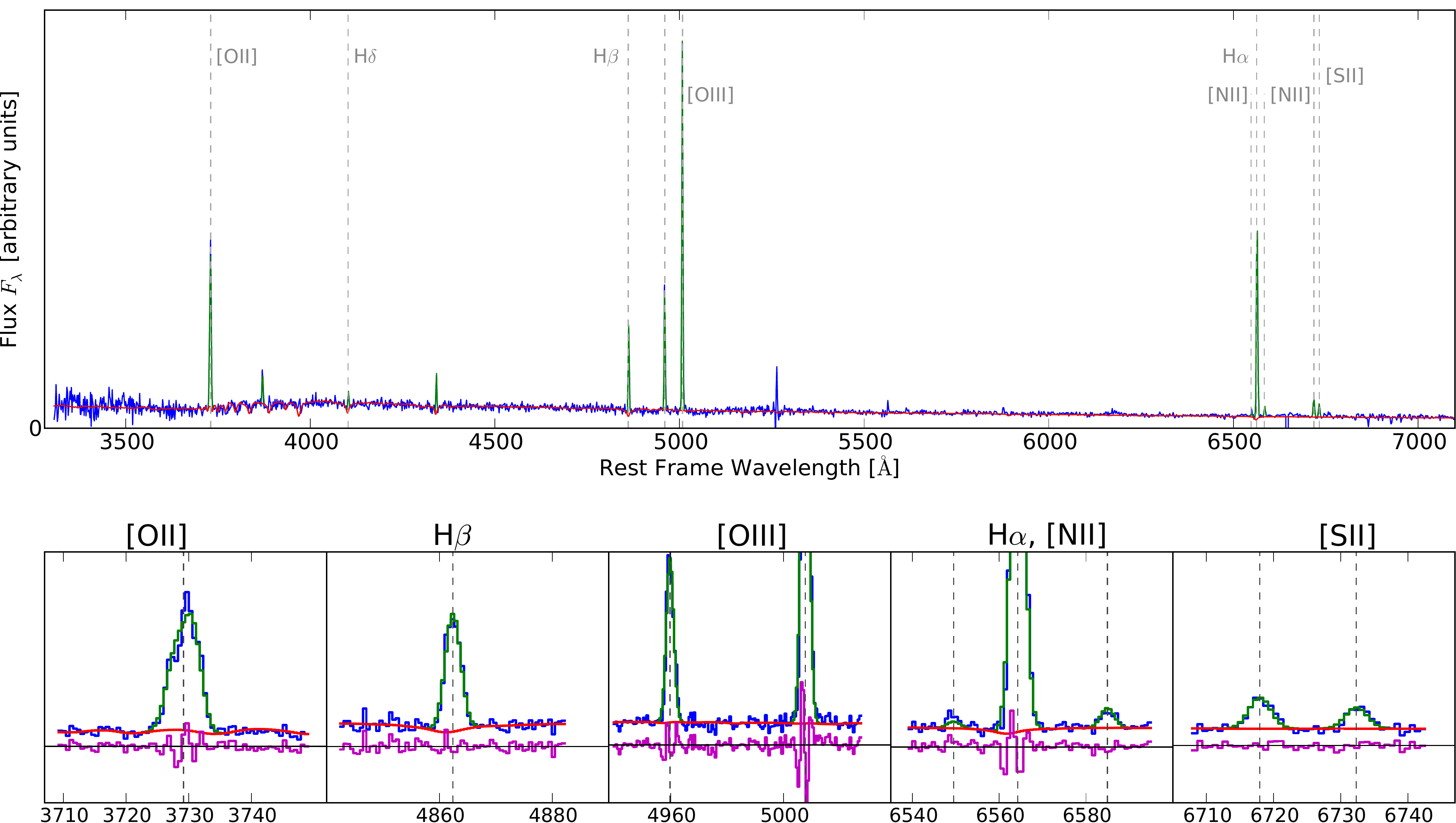}
\end{center}
\caption[Example Spectrum Fit]{An example of our fits to galaxy spectroscopy, here for the host of SNF20070331-013 which was observed with LRIS on Keck. The blue curve is the data, the green is the stellar continuum fit, and the red is the fitted emission line profiles. In the lower panel we show a zoomed-in view of each of the fitted lines, along with the residual difference (magenta curve) between the data and fitted model of the stellar continuum plus emission lines.}
\label{fig:example_emline_fit}
\end{figure*}

\snia\ host galaxy redshifts, metallicities, H$\alpha$ star-formation rates, and internal reddening were calculated using emission line fluxes from the host galaxy spectra.  Accurate measurement of emission line fluxes in star-forming galaxies requires proper accounting for stellar absorption.  To this end we fit the emission line fluxes and stellar background in each host spectrum simultaneously using a modified version of the IDL routine {\tt linebackfit} from the {\tt idlspec2d}\footnote{http://spectro.princeton.edu/idlspec2d\_install.html}  package developed by the SDSS team at Princeton. This routine allows the user to provide a list of template spectra which are then fitted to the data in linear combination with Gaussian emission line profiles.  We have modified this code to force the coefficients multiplying the stellar continuum templates to be non-negative and to fit for reddening of the stellar continuum (using a CCM law with $R_V=3.1$ fixed and $E(B-V)$ as a fit parameter). Additionally, we have incorporated the ability to fit for a scaling factor between the blue and red channels of two-arm spectrograph data. For background templates we chose a set of simple stellar populations (SSPs) from the stellar population synthesis code GALAXEV \citep[][BC03]{bc03} with a \citet{chab03} IMF and the same age sampling used for background fitting by \citet{trem04}, which ultimately consists of ten SSPs for each metallicity track. These templates are convolved to the resolution of the particular spectrograph whose data we fit. We note that the use of \citet{salpeter} IMF templates results in negligible differences to the fitted emission line fluxes, and metallicity difference smaller than our typically quoted precision of 0.01~dex. An example fit to spectroscopic data is shown in Figure~\ref{fig:example_emline_fit}.

Use of the continuum model naturally accounts for the effect of Balmer absorption. Given our typically low spectral resolution (e.g. $~400$~km~s$^{-1}$ at \ha\ and $~200$~km~s$^{-1}$ at \hb\ for Kast), we did not fit for a velocity dispersion for the stellar continuum in each \snia\ host galaxy spectrum since our templates were already broadened by the spectral resolution. This is adequate for nearly all host galaxy and instrument combinations, where the expected galaxy velocity dispersion (based on $\sigma_v\propto M^{1/4}$) is either below the spectrograph resolution or \ha\ is strong enough that stellar absorption corrections are very small relative to the emission strength. For example, we re-fit the Kast spectrum of the strongly star-forming ($EW(H\alpha) = 75$\AA) host of SNF20050826-004 with stellar templates broadened by $~400$~km~s$^{-1}$, and found the \ha\ flux changed by less than 2\%. There may be some massive ellipticals with weak emission (e.g. $\log(M_*/M_\odot) > 11$ and $EW(H\alpha) < 5$\AA) for which velocity dispersion could be important, but these cases typically have emission dominated by AGN activity and ultimately do not produce a gas-phase metallicity.

\begin{figure}
\begin{center}
\includegraphics[width=0.45\textwidth]{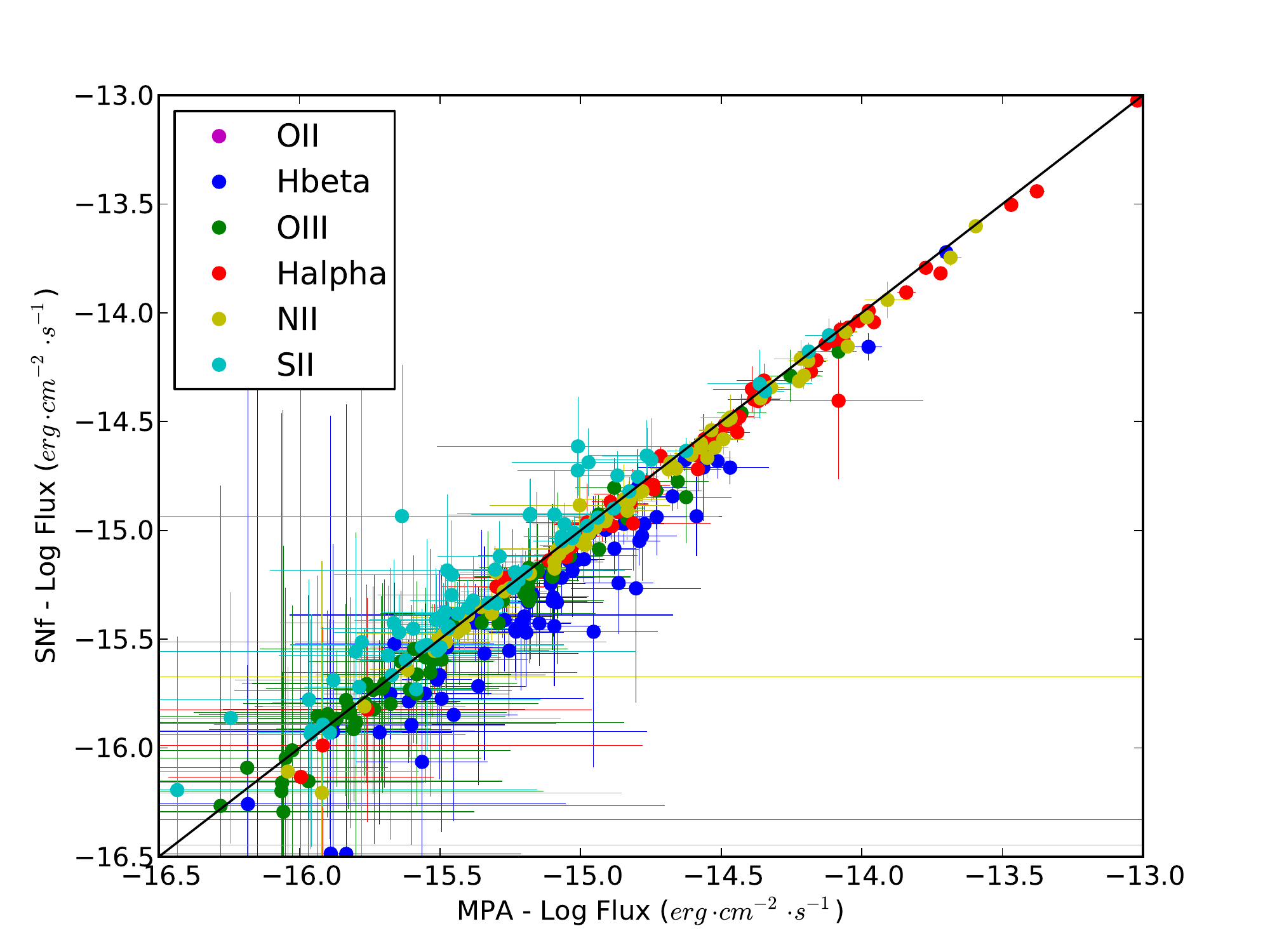}
\end{center}
\caption[Emission Line Fit vs. SDSS]{A comparison of the fitted emission line fluxes from our modified emission line fitting code (y-axis) vs. the values derived by the MPA-JHU team (x-axis). Points are color coded by emission line, and the line representing unity is the solid black line.}
\label{fig:sdss_mjc_emline_compare}
\end{figure}

To ensure that our modifications of the code are not producing anomalous results, we compared our fitted emission line fluxes to those derived by the MPA-JHU team for those hosts whose spectra were obtained from SDSS. In Figure~\ref{fig:sdss_mjc_emline_compare}, we plot these values and show that our results are very consistent with those derived by these authors. Thus we believe our emission line flux estimates, including the resultant corrections for continuum absorption, are accurate. 

The only minor discrepancy may be in the flux of the \hb\ line. Our emission line fits use the BC03 models, whereas the MPA-JHU values for DR7 use the updated (unpublished) Charlot \& Bruzual 2008 models. The BC03 models have been previously shown to underestimate the strength of the \hb\ emission line \citep{groves12} in a comparison between the DR7 and DR4 fits (which used the BC03 models) due to inaccurate fitting of the \hb\ absorption profile. This issue could affect estimates of the galaxy reddening and thus bias metallicity measurements. For example, an underestimate of the \hb\ line would over-estimate the reddening, which would result in an over-estimate of the OII~$\lambda$3727 flux, which in the $R_{23}$ metric would produce an over-estimate of the metallicity. However, we use a reddening-insensitive metallicity indicator (PP04 $N2$ - see Section~\ref{sec:snf_host_metal}) in all but the lowest metallicity \snia\ hosts. The \hb\ bias (of order a few \AA\ in equivalent width) is strongest in galaxies with weak emission, and our low-metallicity hosts are all very strongly star-forming (EW(H$\beta$)$\sim$50\AA) and thus likely to be only marginally affected by this problem.

Redshifts for \snf\ host galaxies with strong emission lines were derived as the weighted (by measurement uncertainties) mean of individual emission line redshifts fitted from host spectra. Redshift errors were similarly calculated from the measurement uncertainties on the individual line redshifts. This method is the same as that used by SDSS for final redshifts reported in their online database. For hosts with very weak or no emission lines, redshifts were calculated with a cross-correlation technique using the methods presented by \citet{tondav79}. We correlated the best fit stellar continuum spectrum against the observed host spectrum after subtraction of the fitted emission line fluxes. Typical redshift errors for these two methods are of the order $\sigma_z \sim 0.0001$, and tend to be limited by the precision of the spectroscopic wavelength solution.

\begin{figure}
\begin{center}
\includegraphics[width=0.45\textwidth]{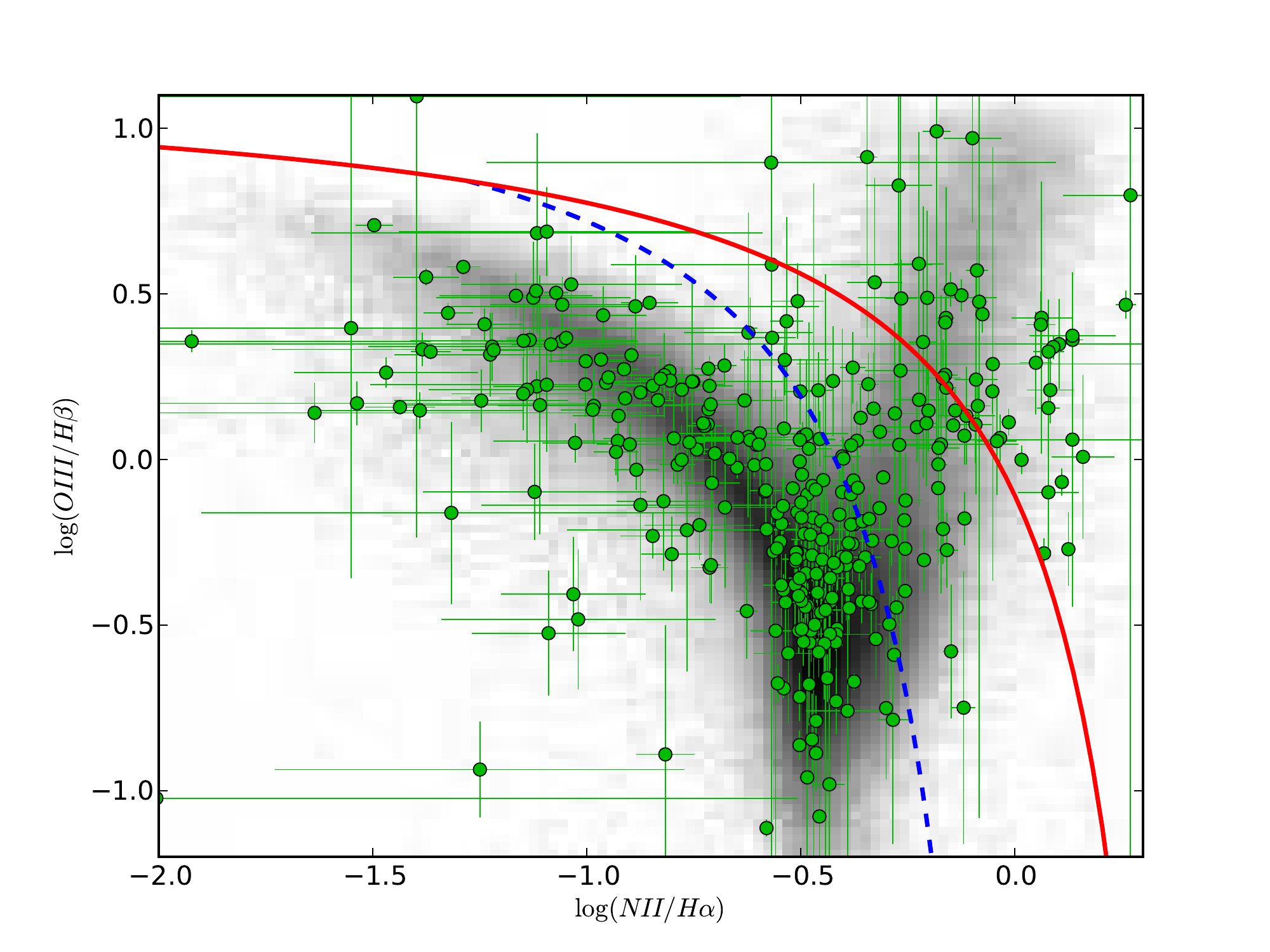}
\end{center}
\caption[\snf\ Hosts BPT Diagram]{\snia\ hosts from \snf\ in the BPT diagram. The diffuse grey background represents SDSS galaxies whose emission line fluxes were measured by the MPA-JHU team. Galaxies above the solid red line are classified as ``AGN'' galaxies, those below the dashed blue line are ``star-forming'' galaxies free of AGN contamination, and those galaxies between the two lines are classified as ``composite'' galaxies.}
\label{fig:snf_bpt}
\end{figure}

In order to derive gas-phase metallicities (Section~\ref{sec:snf_host_metal}) or H$\alpha$-based star-formation rates (Section~\ref{sec:snf_ha_sfr}), we will invoke empirical formulae derived under the assumption that nebular emission is excited by ionizing flux from young stars. To do so we must first excise those galaxies whose emission line fluxes are contaminated by AGN activity using the emission line diagnostic diagram of \citet[][hereafter BPT]{bpt81}. In Figure~\ref{fig:snf_bpt}, we plot the distribution of \snf\ emission line host galaxies on the BPT diagram as compared to the distribution of galaxies from SDSS, with the boundaries defined by \citet{kewley06} to distinguish normal star-forming galaxies from AGNs and composite galaxies. Of the 374 host galaxies with spectra, 315 had good detections of all four lines used in this BPT diagram. Of those, 215 had emission line fluxes consistent with ionization by young stars, 50 had line fluxes indicating AGN ionization, and 50 had line flux ratios in the composite region of the diagram. This high AGN fraction is likely a product of our spectra being dominated by light from the galaxy core where AGN activity is strongest.

Finally, for those hosts whose emission was consistent with star-formation, emission line fluxes were corrected for internal reddening within the host galaxy by employing the Balmer decrement method. In an HII region ionized by young stars, the ratio of emission line flux in the \ha\ line to that in the \hb\ line is fixed by atomic physics. For typical HII regions whose temperatures are around 10,000~K, this ratio is calculated to be $F(H\alpha)/F(H\beta) = 2.87$ under Case B recombination \citep{agn2}, and is known as the Balmer decrement. Reddening by dust causes the observed value of this flux ratio to exceed its canonical value, and one can calculate the amount of reddening by assuming a reddening law such as that of \citet{ccm}. For our \snia\ host galaxies, final emission line fluxes used for calculations of gas-phase metallicity and star-formation rate (from H$\alpha$) have been corrected for the internal reddening calculated using this method. 

\subsubsection{Host Gas Phase Metallicities}
\label{sec:snf_host_metal}
Translating emission line fluxes into a gas-phase metallicity depends on the choice of metallicity calibration, as thoroughly detailed by \citet{ke08}. Different calibrations are known to disagree by as much as $\sim 0.5$ dex, which makes it difficult to place metallicity measurements on a common absolute scale.  Additionally, there is no single metallicity metric that is ideal across the entire range of metallicity probed by our sample.  For example, metrics that rely on the NII $\lambda$6584 line, such as \citet{kd02} and \citet{pp04} methods, have high signal-to-noise at high metallicity and are monotonic, but at low metallicities this line becomes very weak and produces large errors in metallicity measurements. The well-known R23 metric \citep[e.g.][]{m91, z94, kk04} is double-valued with metallicity, and is shallow-sloped (in terms of R23 versus metallicity) at low metallicity (i.e. flux errors propagate into larger metallicity errors).  At low metallicities, the preferred metallicity calibration is the $T_e$(OIII) method \citep{aller84}, which relies on the auroral $\lambda$4363 oxygen line.  This method is considered the most reliable, but relies on a very weak emission line and does not consistently agree with the empirical strong-line methods.

Thus it is challenging to find a consistent metallicity calibration that has high sensitivity over the full observed range of galaxy gas-phase metallicities. Additionally, the lack of consistency of absolute metallicity scales between various calibrations makes it difficult to compare reported metallicity values from numerous authors. In order to utilize the strongest available lines and place our measurements on a well-known common scale, we employ different calibrations at different scales and then place all our metallicities on the common \citet[][hereafter T04]{trem04} scale using the conversion formulae presented in \citet{ke08}. 

For galaxies with $\log(\mathrm{NII}/\mathrm{H}\alpha) > -1.3$ (i.e. ``high'' metallicity galaxies), we use the ``N2'' method of \citet{pp04}, as the NII/H$\alpha$ ratio is a sensitive metallicity indicator in this range and has relatively low sensitivity to reddening due to its short wavelength baseline. For very low metallicity galaxies with $\log(\mathrm{NII}/\mathrm{H}\alpha) < -1.3$, we use the ``R23'' method of \citet{kk04} as updated by \citep{ke08}, as this method depends on the relatively strong oxygen lines and also fits iteratively for the ionization parameter. Although the R23 metric is doubly valued with metallicity, the choice of our $\log(\mathrm{NII}/\mathrm{H}\alpha)$ cut places these galaxies firmly on the low-metallicity ``branch.'' Metallicities calculated from these original methods are finally converted to the T04 scale using the conversion formulae of \citet{ke08}. It is worth noting that the dispersion in these conversion formulae is comparable to the systematic uncertainty in the metallicity calibrations themselves. Thus unless otherwise noted, all metallicities reported here are on the T04 scale after application of the above described conversion. We report the  metallicity method and final T04 metallicity for all our \snia\ hosts in Table~\ref{tab:host_specres}, along with the BPT classification of our host galaxies, and the reddening value calculated from the Balmer decrement method.

\begin{table*}
\begin{center}
\caption{\snf\ \snia\ Host Spectroscopic Results}
\label{tab:host_specres}
\begin{tabular}{lcccccc}
\hline
SN Name & BPT Class & Z Method & $Z_{T04}$ & Balmer $E(B-V)$ & EW(H$\alpha$) & $\log(F(\mathrm{H}\alpha))$ \\
\hline
SNF20050621-001 & SF & N2 (PP04) & $9.11^{+0.04}_{-0.04}$ & $0.058 \pm 0.145$ & 19.51 & 41.34 \\
SNF20050624-000 & AGN & \nodata & \nodata & \nodata & 30.00 & 41.65 \\
SNF20050704-008 & SF & N2 (PP04) & $9.02^{+0.03}_{-0.03}$ & $0.557 \pm 0.082$ & 22.47 & 41.28 \\
SNF20050727-005 & AGN & \nodata & \nodata & \nodata & 10.53 & 41.00 \\
SNF20050728-000 & SF & N2 (PP04) & $8.36^{+0.08}_{-0.10}$ & $0.000 \pm 0.044$ & 47.27 & 40.50 \\
SNF20050728-006 & Comp. & \nodata & \nodata & \nodata & 10.91 & 40.53 \\
SNF20050729-008 & SF & N2 (PP04) & $8.98^{+0.03}_{-0.03}$ & $0.933 \pm 0.151$ & 19.63 & 41.07 \\
SNF20050730-003 & AGN & \nodata & \nodata & \nodata & 1.47 & 40.32 \\
SNF20050731-011 & Comp. & \nodata & \nodata & \nodata & 6.12 & 41.02 \\
SNF20050820-004 & Comp. & \nodata & \nodata & \nodata & 0.97 & 40.41 \\
\hline
\end{tabular}
\normalsize
\end{center}
\end{table*}

\subsubsection{Host Star-Formation Rates from \ha\ Flux}
\label{sec:snf_ha_sfr}
The flux in the \ha\ line is often used to calculate the current star-formation rate (SFR) in galaxies. We calculate \ha-based SFRs for our star-forming (i.e. non-AGN) hosts using the following method. We first measure the \ha\ flux from the spectrum, then correct it for reddening using the Balmer decrement as described above. To determine the total amount of \ha\ flux corrected for spectroscopic slit-loss, we calculate a synthetic $i$-band magnitude from our galaxy spectrum and compare it to the observed $i$-band magnitude to calculate a scaling factor. From this total observed galaxy \ha\ flux, we calculate the total rest-frame \ha\ luminosity using the distance modulus at the galaxy's redshift calculated with the code of \citet{wright06}. Finally we convert this \ha\ luminosity to a SFR using the formula of \citet{kenn98} then multiply by 0.7 to convert the Kennicutt formula from a \citet{salpeter} IMF to our \citet{chab03} IMF.

\begin{figure}
\begin{center}
\includegraphics[width=0.45\textwidth]{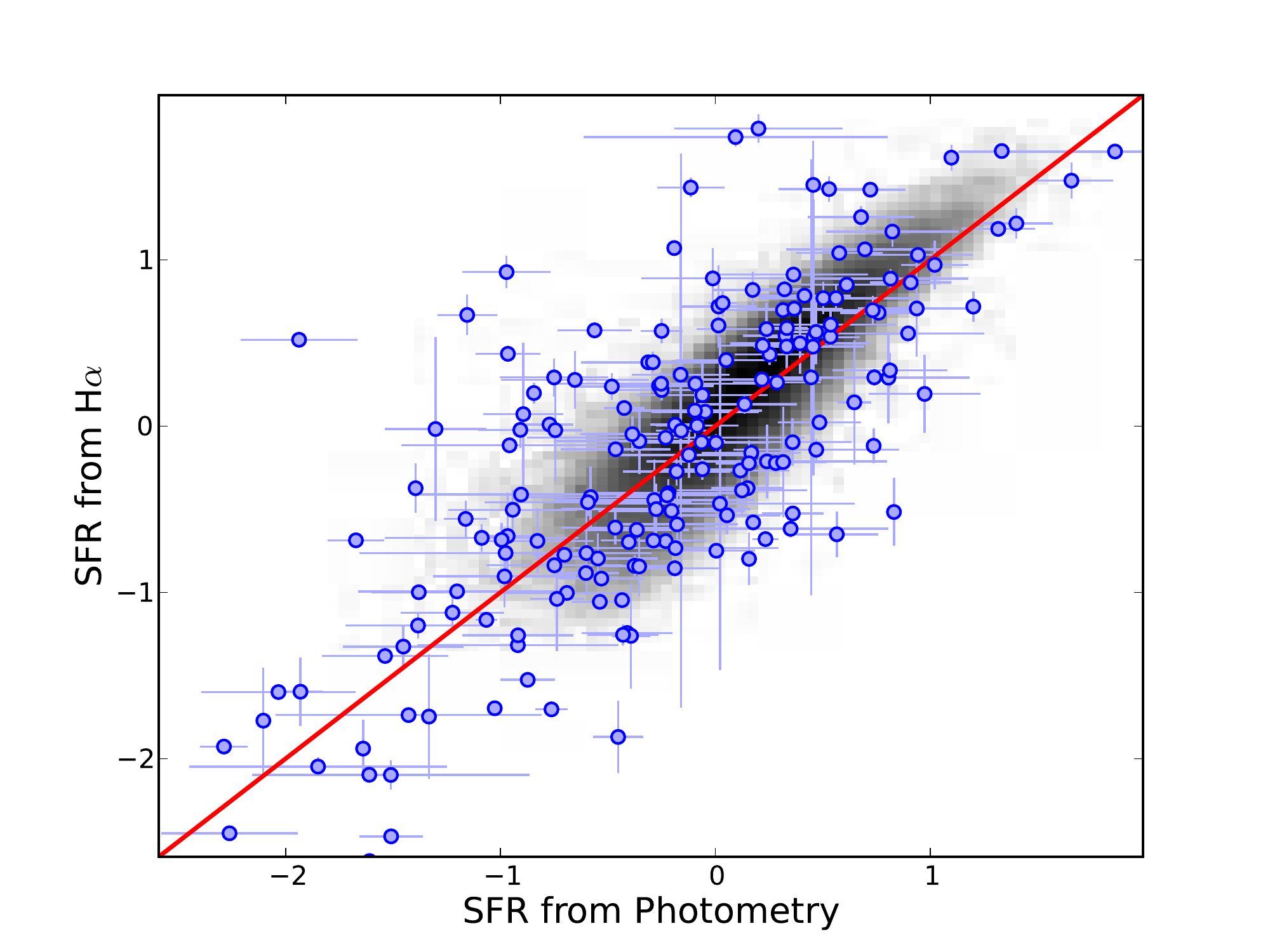}
\end{center}
\caption{Star-formation rates for \snf\ \snia\ hosts (blue circles) calculated from \ha\ flux in spectra versus those calculated from broadband photometry (Section~\ref{sec:host_phot_sps}). The same quantities calculated for normal star-forming galaxies from SDSS are shown as the grey density plot, while a line of unity is shown as the thick red line.}
\label{fig:ha_vs_phot_sfr}
\end{figure}

We can compare these \ha-based SFRs to those calculated from photometry using our SPS methods (see Section~\ref{sec:host_phot_sps}), and plot this comparison in Figure~\ref{fig:ha_vs_phot_sfr}. In the same figure, we plot the SFRs for SDSS galaxies calculated from \ha\ flux measured by the MPA-JHU team coupled to the same \citet{kenn98} formula against SFRs calculated by that group from broadband galaxy photometry \citep{brinchmann04}.  As can readily be seen, the relation between spectroscopic \ha\ SFRs and photometric SFRs is very tight for SDSS, with a dispersion about the unity line of 0.22~dex. For \snf\ hosts the agreement is also good but with an increased scatter of 0.66~dex which is larger than the typical measurement error of 0.23~dex for the \ha-based SFR values. The smaller scatter in the SDSS data is likely due to the fact that the  MPA-JHU SFR values are the sum of the \ha-based SFR for galaxy flux in the SDSS spectroscopic fiber and a color-based SFR estimate for galaxy flux outside the fiber trained on fiber colors and \ha\ fluxes for the full spectroscopic sample. A more detailed discussion of these effects will be given in Section~\ref{sec:snia_host_sfr}. We note that the SFRs of the \snf\ sample extend much lower than the nominal SDSS sample plotted here, due to the greater representation of low mass galaxies in the \snf\ host sample.

\section{Host Galaxy Masses and Star-Formation Rates from Photometry}
\label{sec:host_phot_sps}
Calculation of a galaxy's stellar mass and star formation rate from its photometry requires the use of stellar population synthesis (SPS) techniques. The core principle of this technique involves using model stellar population spectral energy distributions (SEDs) to predict the flux in various photometric filters, then comparing these model predictions to observations. SPS techniques typically combine model SEDs for stars of a single age with masses distributed according to some initial mass function (IMF), thereby deriving the SED for what is known as a simple stellar population (SSP) of uniform age and metallicity. Full galaxy SEDs are calculated by preparing a model star-formation history (SFH) and convolving the SSP SEDs with the relative weights prescribed by the galaxy SFH.

The field of galaxy stellar population synthesis is a rich and constantly evolving field. The best SPS models require stellar evolutionary tracks as well as observed (and/or modeled) stellar SEDs spanning the full parameter space of stellar properties. While most SPS techniques give qualitatively similar results, it is important to understand and track the differences between SPS techniques employed by different authors. Perhaps the two most popular sets of models in the past decade have been GALAXEV \citep[][hereafter BC03]{bc03} and PEGASE \citep{pegase}. Most of the major analyses of the SDSS galaxy sample employed the BC03 models \citep{kauff03a, brinchmann04, trem04, salim07}. Many \snia\ host galaxy studies from recent years \citep{sullivan06, howell09, neill09, kelly10, sullivan10, lampeitl10} have made use of PEGASE, but fortunately the different SPS models give consistent results when scaled appropriately \citep[see, e.g.,][]{kelly10}.  These particular \snia\ host studies employed the code \zpeg\ \citep{zpeg}, which matches observed photometry to PEGASE models for a set of galaxy evolutionary scenarios input by the user [though \citet{gupta11} use the Flexible SPS code of \citet{fsps} and a probabilistic galaxy parameter estimation approach in some ways similar to ours]. Though designed primarily as a tool for deriving photometric galaxy redshifts, \zpeg\ inherently derives galaxy masses and star-formation rates by choosing a best model SED and scaling it to the observed galaxy photometry.

For our purposes we desire not only an accurate estimate of our host galaxy properties (stellar mass, SFR, reddening) but also a proper estimation of how well the data constrain those properties (i.e. their uncertainties). This requires that we calculate not only how well a single model galaxy SED is fitted by our observed photometry, but also what range of models can fit our data and what this implies for the allowable range of galaxy properties. This inherently requires that we employ a Bayesian approach, which we will describe in detail in the following section.

\subsection{Bayesian Stellar Population Synthesis of \snia\ Host Galaxy Photometry}
Deriving the physical properties of a galaxy from broadband photometry is a challenging task. Galaxies are composed of billions of stars of various masses and ages, embedded within some nontrivial geometric distribution of gas and dust. A small discrete set of observable quantities, namely broadband photometry, give insufficient degrees of freedom to independently constrain the full stellar population and dust distribution within a galaxy. Instead we must establish some reasonable prior constraint on how the ages of stars within that galaxy are likely to be distributed by positing the functional form of the galaxy star-formation history (along with a similar prior on dust), as well as the mass distribution produced by each generation of stars (the IMF). With this SFH prior and a chosen IMF we then use the observational data for a particular galaxy to infer the most likely star-formation history as well as its uncertainty.
In practice, we implement this Bayesian inference technique by using SFH and dust priors to produce a large sample of Monte-Carlo generated SFHs, which then define model SEDs when coupled to the SPS models of BC03. We then constrain three key properties of our observed galaxies -- mass-to-light ratio, specific star-formation rate (sSFR), and dust reddening -- by summing over the values of those properties for each model galaxy, weighting by how well the model photometry matches the observed photometry.

Our method is inspired by, and closely resembles, the methodology employed by \citet{kauff03a, gallazzi05, salim07, gb09}. We randomly generate 150,000 model galaxy SFHs as follows. Model galaxy SFHs consist of an exponentially declining star-formation rate (SFR~$\propto \exp(-\gamma t)$) with the decay constant ($\gamma$) distributed uniformly between 0 and 1~Gyr$^{-1}$ and star-formation beginning at a time $t_\mathrm{form}$ distributed uniformly between 1.5 and 13.5~Gyr in the past. In addition to the exponential continuous star-formation component, each SFH is augmented with random bursts of star formation whose durations are distributed uniformly in the range 30~Myr$\leq t_\mathrm{burst} \leq$300~Myr, with mass distributed logarithmically between 0.3 and 4.0 times the total mass formed in the continuous SF component. These bursts of star formation occur with a random likelihood of occurrence such that 50\% of model galaxies have experienced a burst of SF in the past 2~Gyr. Each model galaxy is assigned a single metallicity (i.e. no chemical evolution is modeled), with all model metallicities distributed uniformly between $0.2 \leq Z/Z_\odot \leq 2.5$ and as a smoothly decaying function ($\propto\log(Z)^{1/3}$) in the range $0.02 \leq Z/Z_\odot \leq 0.2$ (in order to not over-represent low metallicity galaxies). All model galaxy parameters were generated randomly, and final model galaxy spectra were constructed from the BC03 simple stellar population (SSP) spectra using the \citet{chab03} IMF and the Padova 1994 evolutionary tracks.

Each model spectrum was reddened to simulate dust in the galaxy using the \citet{cf00} reddening law with reddening $E(B-V)$ drawn from an empirical distribution. This empirical distribution was derived from Balmer decrement \citep[e.g.][]{agn2} measurements of star-forming galaxies in the SDSS spectroscopic sample in the redshift range $0.04 < z < 0.10$ as calculated from the emission line fluxes measured by the MPA-JHU team. For galaxies with Balmer decrement equal to or below the canonical value a reddening value of $E(B-V)=0$ was imposed to forbid non-physical negative reddening. This distribution effectively acts as a prior on the expected amount of reddening in normal galaxies, and is conceptually similar to the one employed by \citet{salim07}.

We also inspected the use of an effectively flat prior on reddening by allowing each model to be fitted for an optimal reddening to match observed photometry values. Doing so we found that very blue galaxies with large observational error bars were frequently fitted with non-physical negative reddening applied to the oldest model SFHs. Similarly for red galaxies, each template was fitted with a large amount of reddening so that the reddening (and sSFR) estimates were skewed to a distribution driven by the difference between the model galaxy color distribution and the observed galaxy color. Thus we believe this empirical reddening prior is justified in order to obtain realistic estimates of reddening in our \snia\ host galaxies.

\begin{figure}
\begin{center}
\includegraphics[width=0.45\textwidth]{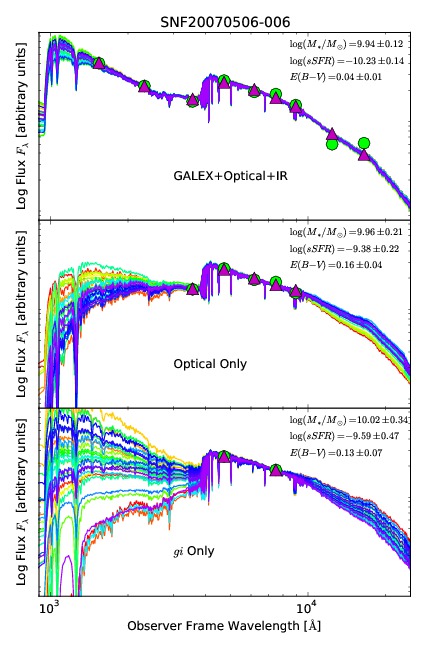}
\end{center}
\caption{Example SPS photometry fit for the host of SNF20070506-006 using three filter subsets: (top) all UV/optical/NIR filters, (middle) optical only, and (bottom) only the $g$ and $i$ filters.  In each panel, the observed photometry points are the large green circles, while the purple triangles are the photometry points for the single best fitting model SED.  The 30 best fitting SEDs are shown, color coded so that purple colors indicate stronger matches while red colors indicate poorer matches.}
\label{fig:bc03_fit_example}
\end{figure}

Galaxy stellar mass, SFR, and reddening were computed for each \snia\ host galaxy as follows. Fluxes for our models were synthesized from their SEDs using the filter throughputs for \galex, SDSS, and 2MASS at several steps in redshift ($0 < z < 0.2$ in steps of $\Delta z=0.005$), then adjusted according to their assigned reddening. For each \snia\ host, we calculate the fluxes of each model SED at the host redshift by interpolating the fluxes computed at the two nearest redshift steps (we note that our choice of fine redshift step resulted in negligible interpolation error of $\leq0.005$~mag). Each model $i$ was then fitted for a simple scaling factor $A_i$ to minimize the $\chi^2$ of its fluxes $F_{\lambda,i}$ matching to the observed photometry $F_{\lambda,obs}$ 
\begin{equation}
  \chi^2_i = \sum_\lambda \left[\frac{F_{\lambda,i}-F_{\lambda,obs}}{\sigma_{\lambda,obs}}\right]^2
\end{equation}
We note here that photometric errors $\sigma_{\lambda,obs}$ for \snia\ host galaxies were padded to ensure physically meaningful $\chi^2$ values, as we will justify in detail in Section~\ref{sec:sps_comparisons}. An example of this fitting procedure is shown in Figure~\ref{fig:bc03_fit_example}.

Each \snia\ host galaxy property (mass-to-light ratio, sSFR, reddening) and its uncertainty was then calculated from the probability distribution function (PDF) constructed as the $\chi^2$ weighted sum of that property  across all SPS models:
\begin{equation}
  \left<p\right> = \sum_i e^{-\chi_i^2/2}p_i
\end{equation}
where the three properties calculated are mass-to-light ratio, sSFR, and reddening. In practice we use the $\chi^2$ weights for each model to calculate the weighted cumulative probability distribution for each quantity, then from this calculate median and $\pm1\sigma$ values for the desired quantity.

Galaxy masses were computed using the observed $g$-band absolute magnitude, corrected for the amount of dimming $A_g$ corresponding to the fitted reddening $E(B-V)$, and the mass-to-light ratio ($M_*/L$) and its error as calculated from the $M_*/L$ PDF. Model specific SFR (sSFR $\equiv$ the SFR per unit mass) values were calculated as the average over the last 0.5~Gyr for each model, and the host galaxy fitted sSFR was calculated from the sSFR PDF. Similarly, the best fit extinction $E(B-V)$ for each host was calculated from the PDF constructed from the input model extinction values.

We explored the possibility of using $i$-band mass-to-light ratios instead of $g$-band, since this band has formally smaller scatter in mass-to-light ratio \citep[see e.g.][]{gb09}. We found that for our faintest host galaxies $i$-band yielded a much larger mass error due to the blue colors of these galaxies and increased sky brightness in $i$-band yielding a larger photometric error than $g$-band. For consistency then we chose to use $g$-band for mass estimates for all our \snia\ host galaxies. We found that the host galaxy mass values derived from $i$-band were consistent with our nominal $g$-band mass-to-light ratio values, but with larger uncertainties.

In Table~\ref{tab:host_sps_res} we present the fitted mass, sSFR and host reddening $E(B-V)$ fitted for our \snf\ hosts using our SPS methods. In Figure~\ref{fig:bc03_fit_example} we plot a representative example fit to photometry for one of our \snf\ host galaxies, along with photometry fits using reduced sets of filters. The ability of our photometry to constrain host galaxy physical parameters was highly dependent on the filter sets used, with significant improvement on the sSFR and extinction measurements when \galex\ data was included. We discuss the implications of these effects in the following section.

\begin{table*}
\begin{center}
\scriptsize
\caption{\snf\ \snia\ Host Galaxy Stellar Population Properties}
\label{tab:host_sps_res}
\begin{tabular}{lccc}
\hline
SN Name  & $\log(M_*/M_\odot)$ & $sSFR$ & Phot. E(B-V) \\
\hline
SNF20050519-000 & $10.55^{+0.48}_{-0.36}$ & $-10.21^{+0.40}_{-0.52}$ & $0.48^{+0.19}_{-0.25}$\\
SNF20050621-001 & $10.45^{+0.25}_{-0.25}$ & $-11.10^{+0.40}_{-0.29}$ & $0.10^{+0.15}_{-0.10}$\\
SNF20050624-000 & $10.73^{+0.13}_{-0.14}$ & $-10.77^{+0.14}_{-0.14}$ & $0.00^{+0.06}_{-0.00}$\\
SNF20050704-008 & $10.28^{+0.01}_{-0.01}$ & $-9.08^{+0.00}_{-0.00}$ & $0.35^{+0.00}_{-0.00}$\\
SNF20050727-005 & $10.32^{+0.09}_{-0.16}$ & $-10.85^{+0.15}_{-0.18}$ & $0.00^{+0.08}_{-0.00}$\\
SNF20050728-000 & $8.91^{+0.20}_{-0.19}$ & $-9.27^{+0.31}_{-0.18}$ & $0.18^{+0.07}_{-0.09}$\\
SNF20050728-001 & $10.08^{+0.11}_{-0.14}$ & $-13.63^{+0.30}_{-0.60}$ & $0.00^{+0.05}_{-0.00}$\\
SNF20050728-006 & $10.23^{+0.19}_{-0.21}$ & $-10.08^{+0.18}_{-0.20}$ & $0.43^{+0.10}_{-0.09}$\\
SNF20050728-012 & $11.14^{+0.22}_{-0.25}$ & $-10.13^{+0.33}_{-0.30}$ & $0.59^{+0.13}_{-0.12}$\\
SNF20050729-008 & $10.38^{+0.15}_{-0.16}$ & $-10.39^{+0.30}_{-0.37}$ & $0.10^{+0.07}_{-0.10}$\\
\hline
\end{tabular}
\normalsize
\end{center}
\end{table*}

\subsection{The Effect of Uneven Photometric Coverage}
\label{sec:filter_coverage}
Because our host galaxies do not all have the same set of photometric filters, it is vital to ensure that this uneven coverage does not bias our results. To this end, we tested how well our method was able to recover the input model physical parameters when restricted to photometric matching with a reduced set of filters. Because a few of our \snia\ hosts were not imaged by \galex\ (generally due to proximity to bright stars) or were too faint to have significant detections in \galex\ or 2MASS, we choose to examine the effect of having only optical $ugriz$ photometry on galaxy parameter recovery. Similarly, since our SNIFS observations of faint hosts specifically target $g$ and $i$ band photometry, we also examined how well galaxy properties could be recovered with only this single color.

\begin{figure*}
\begin{center}
\includegraphics[width=0.90\textwidth]{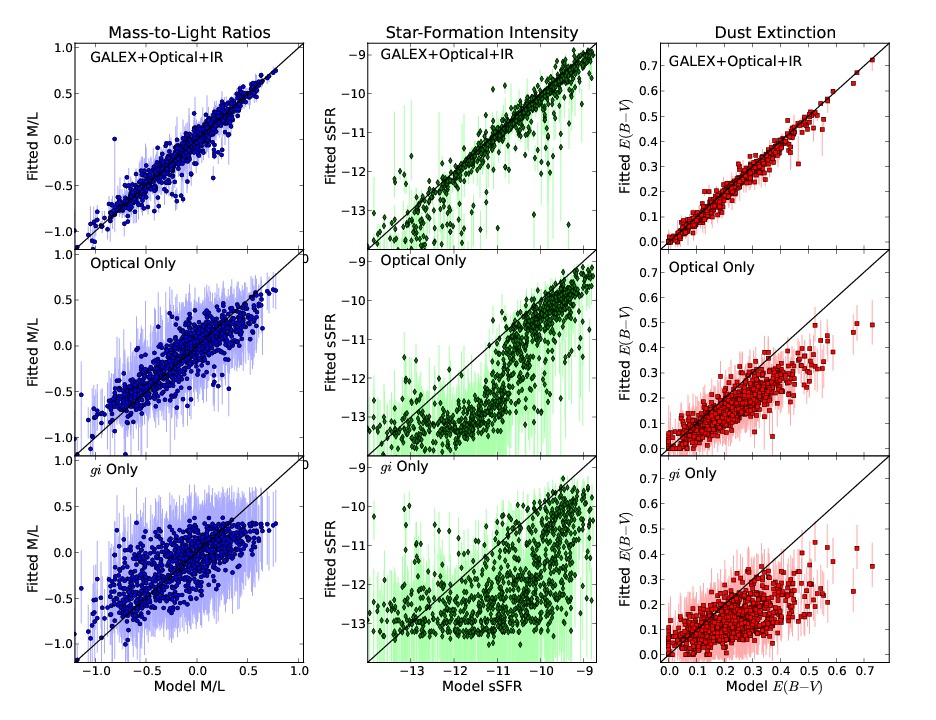}
\end{center}
\caption{Model parameter recovery efficiency for a sample of 1000 model galaxies using different subsets of photometric filters. The top row shows the recovery of model mass-to-light ratios, specific star formation rates, and reddening by dust when using all filters from the UV to NIR. The middle row shows the same results using only optical filters ($ugriz$), while the bottom row shows the recovery efficiency for only two optical filters ($g$ and $i$). Mass-to-light ratios M/L are in solar units, sSFR values are $\log(\mathrm{yr}^{-1})$, and extinction values $E(B-V)$ are in magnitudes. Parameter recovery efficiencies are summarized in Table~\ref{tab:param_recov}.}
\label{fig:sps_results_compare}
\end{figure*}

We selected a random sample of 1000 model galaxies and perturbed their tabulated magnitudes by values distributed normally about zero with a dispersion equal to the systematic error values used to pad our observational photometric data (see Section~\ref{sec:sps_comparisons}). We then applied our Bayesian parameter recovery test (excluding each test model from the set of models used in the comparison) using different subsets of filters: (i) all available filters, (ii) optical filters only (no UV/NIR), and (iii) $g$- and $i$-band only. We show in Figure~\ref{fig:sps_results_compare} the comparison of the recovered mass-to-light ratio, sSFR, and reddening values as a function of their input values for each filter subset. A representative example of the effects of reduced filter sets is also shown for real data in Figure~\ref{fig:bc03_fit_example}. A summary of the mean offset (bias) and RMS (dispersion) about the correct input values for each parameter for each filter subset is presented in Table~\ref{tab:param_recov}.

\begin{table}
\begin{center}
\caption{Model Parameter Recovery Efficiencies}
\label{tab:param_recov}
\begin{tabular}{ccr}
\hline
\hline
Parameter & Filter Set & Recov. Eff. \\
\hline
              & UVOIR   & $-0.005 \pm 0.095$ \\
$\log(M_*/L)$ & Optical & $ 0.014 \pm 0.176$ \\
              & $g$/$i$ & $ 0.024 \pm 0.265$ \\
\hline
              & UVOIR   & $-0.22 \pm 0.49$ \\
$\log(sSFR)$  & Optical & $-0.79 \pm 0.76$ \\
              & $g$/$i$ & $-1.04 \pm 1.16$ \\
\hline
              & UVOIR   & $-0.012 \pm 0.016$ \\
$E(B-V)$      & Optical & $-0.065 \pm 0.053$ \\
              & $g$/$i$ & $-0.087 \pm 0.085$ \\
\hline
\hline
\end{tabular}
\end{center}
\end{table}

Several important trends are evident in the results of this analysis. First, galaxy mass-to-light ratios are recovered with very little bias ($\lesssim 0.02$~dex) regardless of the filter set used. This result illustrates the fact that galaxy mass-to-light ratios are strongly correlated with color, which is set primarily by the mean stellar age of the galaxy and only to higher order by the subtleties of the galaxy's star-formation history. 
Furthermore, the color-luminosity change due to foreground extinction is only slightly dissimilar to the color-luminosity change that accompanies aging in stellar populations. This degeneracy is troublesome for accurately estimating the extinction and sSFR of a galaxy, but is fortuitous in the recovery of mass-to-light ratios from color. Naturally the dispersion in mass-to-light ratio recovery increases as the number of filters decrease. Table~\ref{tab:param_recov} shows this increase, from 0.10~dex for the full UVOIR filter set, to 0.18~dex for optical only, to 0.27~dex for $gi$ only. These dispersions are satisfyingly small compared to both the range of mass-to-light ratios probed by our models ($\sim$1.3~dex) and the full range of galaxies masses ($\sim$5~dex) probed by our \snia\ host galaxy sample.

The second major result for this analysis is that the lack of UV data can severely bias estimates of galaxy star-formation intensities (sSFRs) and extinction from dust. This effect is particularly acute for galaxies with low  star-formation intensity (i.e. sSFR $\leq -10.5$), where the degeneracies between stellar age and extinction become difficult to disentangle. A galaxy with intermediate star-formation intensity and modest reddening (e.g. sSFR $\sim -11$ and $E(B-V)\sim 0.5$~mag) has remarkably similar optical colors to an unreddened older stellar population. Thus galaxies in this sSFR range tend to have both their sSFR and extinction levels underestimated. However, galaxies with strong star-formation (sSFR $\geq -10.5$) generally have successfully recovered sSFRs and extinction, due to the fact that their blue colors cannot be mimicked by anything other than young stars.

As a final remark, we reiterate that these tests were done using members of a galaxy model library whose parent distribution \emph{by construction} was the same as the model whose properties were being measured. Specifically, the reddening for each model was drawn from the same prior distribution it was fitted against, and with the same extinction law used as input to the model library. Also the simulated measurement errors were distributed exactly according to the values used as systematic errors in our model parameter fits. Thus the results here represent the best case scenario in which we fully understand the underlying distribution of galaxy SFHs, the reddening law of the observed galaxies, the distribution of extinction values of those galaxies, and the systematic uncertainties of our photometric measurements. These results then can be considered a systematics floor with regard to our ability to recover galaxy physical properties from photometry using our SPS method.

\subsection{Validation of SPS Methods}
\label{sec:sps_comparisons}
To ensure the \snia\ host galaxy properties derived with our methods are consistent with those derived by other authors, we here examine the results of applying numerous SPS methods to the same photometric data as a means of validating our method. In order to not bias the results of our \snia\ host galaxy property estimation, we conduct our analysis on a training sample of field galaxies with photometric data from \galex, SDSS, and 2MASS.

For simplicity (and limitation of computation time) we searched the SDSS online database in a small patch of sky between $-1^\circ \leq \delta \leq 1^\circ$ and $0^\circ \leq \alpha \leq 45^\circ$ (i.e. in Stripe 82) for galaxies with spectroscopic observations in a redshift range $0.04 \leq z \leq 0.10$. We then similarly searched the \galex\ photometric catalog for objects with detections in both the FUV and NUV filters in the same region of sky. Next we searched the 2MASS Extended Source Catalog (XSC) in the same sky region. Finally we matched the three catalogs together, requiring objects be detected with 5\arcsec\ (i.e. the \galex\ resolution) of the SDSS astrometric position, compiling a total of 3673 galaxies with reliable detections in all 10 photometric filters. This set served as our training sample for refining our galaxy property estimation method, as well as a base sample for comparing the results of employing different galaxy SPS methods.

We note that although these catalog magnitudes do not employ matched apertures in the manner of our \snia\ host galaxy photometry (see Section~\ref{sec:snf_host_phot}), they are the same magnitudes used in the calculation of galaxy physical properties by the MPA-JHU group whose values will serve to validate the accuracy of our SPS methods. While the advantage of matched photometric apertures in galaxy stellar population modeling was not explicitly investigated for this work, we found the matched aperture method especially advantageous for placing upper limits on the UV flux in galaxies with low star-formation. The validation sample of SDSS galaxies employed here has galaxies spanning the full range of star-formation intensity realized in our SPS models, so the choice to use catalog photometry has not excised low-SFR galaxies from our validation sample.

Because our galaxy property (and uncertainty) estimation relies on appropriate weighting of the models in our library, it is vital that our photometric uncertainty estimates be accurate. We inspected this by examining the distribution of best $\chi^2$ values for our training sample of SDSS galaxies and calculating its agreement with the expected distribution of this quantity. We find that use of the nominal catalog photometric errors results in extremely large $\chi^2$ values. We thus add extra systematic errors in quadrature with the formal photometric errors of our measurements as follows: 0.052 and 0.026~mag for \galex\ FUV and NUV filters, respectively \citep[c.f.][]{salim07}; 0.05, 0.02, 0.02, 0.02, and 0.03~mag for SDSS $ugriz$ filters \citep[c.f.][]{blanton07}; and 0.05~mag for each of the 2MASS and UKIDSS $JHK$ filters. The distribution of best $\chi^2$ values using these padded photometric error bars is in better agreement with the expected $\chi^2$ distribution, so we employ them in the galaxy property estimation for our \snia\ host galaxies.

We note that the final $\chi^2$ values are still slightly higher than predicted by the $\chi^2$ distribution, indicating some residual disagreement between our models and the data. This is likely due to the fact that the true underlying distributions of galaxy SFHs, reddening laws, and extinction values are not fully known quantities. Similarly, the best stellar evolutionary tracks, IMF(s), and even stellar SEDs may still require refinement. While these quantities are of interest for proper galaxy modeling, their study is the subject of a large and rich field of inquiry and thus must be beyond the scope of this work. Our choices of SFH, reddening, and metallicity priors as well as IMF, stellar evolutionary track, and systematic photometric uncertainties, are all based on reasonable assumptions which have even been employed in previous studies. Ultimately our models show acceptable agreement with real data and thus we consider them to be good models for comparing to data.

\begin{figure}
\begin{center}
\includegraphics[width=0.45\textwidth]{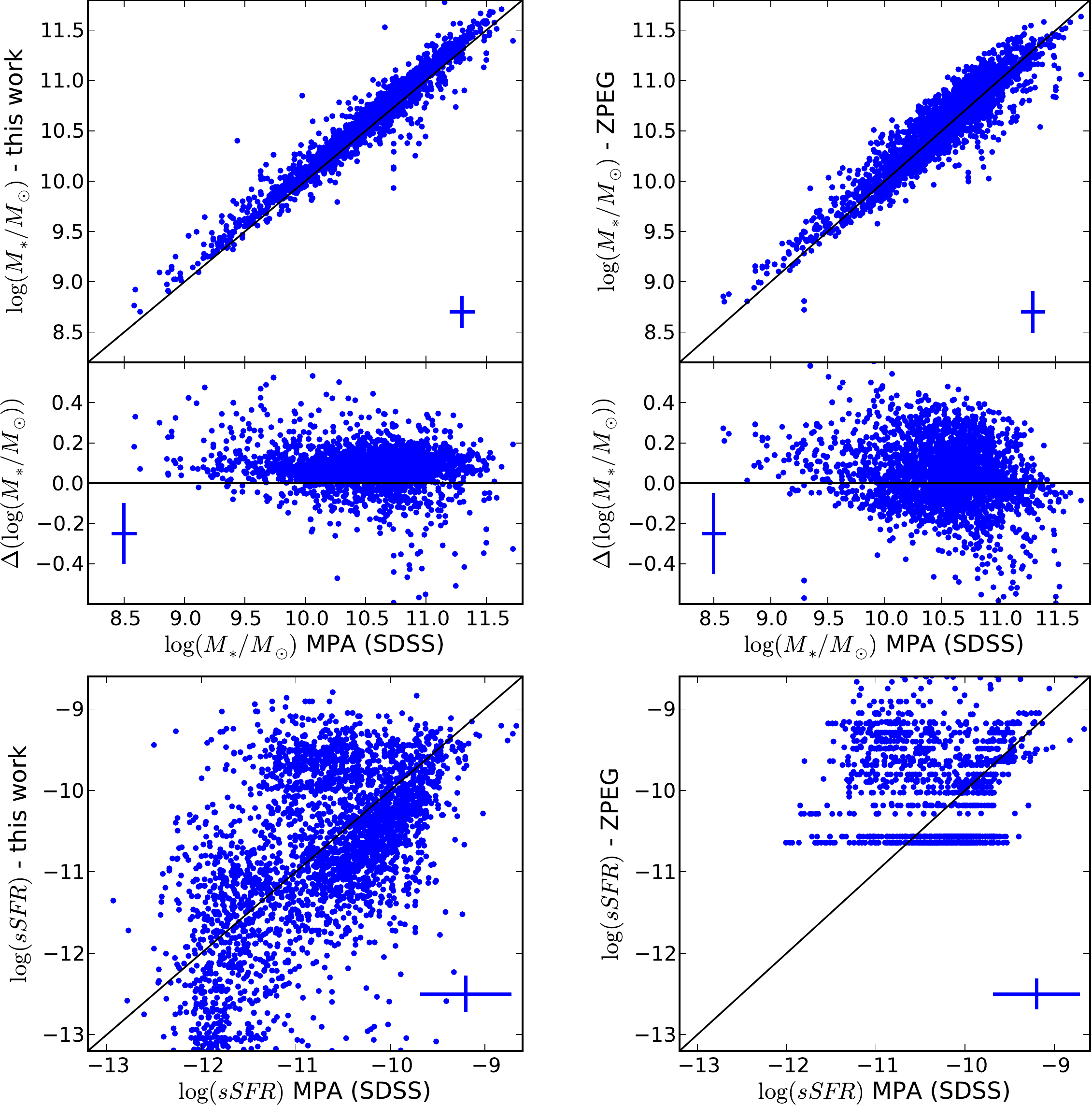}
\end{center}
\caption{Left Panels: Comparison of galaxy stellar masses (top) and specific star-formation rates (bottom) derived using the Bayesian SPS method described here versus the values computed by the MPA-JHU SDSS team for a sample of galaxies in SDSS.  Right Panels: Same as left, but comparing the \zpeg\ values to those of the MPA-JHU group.}
\label{fig:mpa_zpeg_compare}
\end{figure}

To confirm that our Bayesian method is recovering the correct galaxy physical properties, we compare the values we derived for our training sample of SDSS galaxies to the values derived by previous authors. To this end we use the stellar mass and SFR values calculated by the MPA-JHU group for these same galaxies. In the left hand panels of Figure~\ref{fig:mpa_zpeg_compare} we show the values we derive compared to their values. We find our stellar masses to have excellent agreement, with a mean and RMS difference of $\Delta\log(M) = 0.063\pm0.105$ and typical galaxy mass errors of $\sigma_{\log(M)}=0.16$. Their values use the same SPS models (BC03) as us with the same IMF \citep{chab03}, as well as model galaxy SFH parameters drawn from a similarly constructed prior and a similar Bayesian weighting scheme, so such tight agreement is to be expected. 

Our SFR values show a bit more dispersion compared to the MPA-JHU values, with $\Delta\log(SFR) = -0.4\pm1.2$. This is not unexpected for several reasons. The first and most important is that the MPA-JHU SFR values are based on a combination of the SFR calculated from \ha\ flux within the SDSS fiber and color-based SFR measurement for light outside the fiber, which is trained on the \ha\ fluxes and colors of fiber spectra. The second cause for SFR discrepancies is the fact that the \ha-trained SFR values for the SDSS sample are sensitive to the most recent star-formation and may not correlate exactly with our UV-driven SFR values averaged over the last 500~Myr of the galaxy SFH. These effects will be discussed in detail in Section~\ref{sec:snia_host_sfr}.

Next we compute mass and SFR estimates for our training sample of galaxies using the code \zpeg\ \citep{zpeg}. This code was employed by several recent authors \citep{kelly10,sullivan10,lampeitl10} to examine trends in stretch- and color-corrected \snia\ Hubble residuals with the masses of their host galaxies, and thus we want to ensure our values are on a consistent scale with theirs. We used our catalog photometry for our training sample of galaxies and computed masses and SFRs using \zpeg\ with the standard set of templates included with the \zpeg\ code. The comparison of the values derived by the MPA-JHU groups and those derived with \zpeg\ is presented in the right panels of Figure~\ref{fig:mpa_zpeg_compare}, which shows that the stellar mass values derived with \zpeg\ are highly consistent with those derived by the independent analysis of \citet{kauff03a}. The comparison of the stellar mass values from \zpeg\ with our own values also showed similar agreement. We note however that the discrete set of input galaxy evolutionary scenarios in \zpeg\ results in nearly discrete values of fitted sSFR.

We found that the formal \zpeg\ mass and sSFR error bars are significantly smaller than the values derived with our method. This is because \zpeg\ calculates errors on these quantities based on how well the input photometry matches the single evolutionary scenario that fits best, but has no prescription for how uncertain the galaxy SFH might be. Our method implicitly incorporates this estimation by performing a weighted sum over a wide range of galaxy SFHs, so we believe that our error bars more accurately reflect the systematic errors in the galaxy mass and SFR estimation
\citep[see also][for a thorough discussion on the superior results achieved with Bayesian galaxy parameter measurements compared to using the ``best-fit'' template from a discrete set of models]{taylor11}.

Thus we conclude that our mass values are appropriate for use in \snia\ cosmological analyses and are consistent with other methods including \zpeg\ and \citet{kauff03a}. Indeed, since the central values of all these methods are so consistent, it is highly unlikely that trends of \snia\ Hubble residuals with host galaxy mass can be measured to higher precision by employing more sophisticated SPS methods. However, if these host galaxy mass values should ultimately be used as \snia\ luminosity correction parameters, then the estimation of their error bars must be accurate in order to be properly accounted for in the \snia\ cosmology error budget. Thus we propose that our Bayesian method is more realistic as it accounts for galaxy modeling systematics where some other methods do not.

\subsection{The Importance of \galex\ UV Photometry}
\label{sec:uv_limits}
We further investigated the importance of including UV photometry from \galex\ by repeating our galaxy photometry fits without UV data, both for the training sample of SDSS galaxies and our sample of \snia\ host galaxies. In particular we focused on the change in the estimated galaxy sSFR when UV data was not included in the photometry fits.

For the sample of $\sim 3700$ SDSS galaxies in our training sample, we found that the sSFR values calculated from fitting only optical data were on average 0.22~dex lower than the values obtained when utilizing GALEX data. This is in good agreement with Figure~\ref{fig:sps_results_compare}, where we found that the lack of UV data resulted in many intermediate sSFR galaxies with modest dust being mistaken for low sSFR galaxies with no dust. Furthermore, the mean uncertainty in the estimation of sSFRs went up from 0.2~dex to 0.72~dex when UV data was not included in the SPS fit. Thus we see that UV data is critical for both obtaining the correct value of a galaxy's star formation intensity as well as reducing its measurement uncertainty.

We also specifically investigated the effect of neglecting UV data when fitting the photometry of massive elliptical (passive) galaxies. To investigate this portion of parameter space, we isolated the subset of $\sim 950$ galaxies in our training sample with $\log(M_*/M_\odot) > 10.5$ and $\log(sSFR) < -11.5$. For these galaxies we found the sSFR values fitted from only optical photometry were 0.28~dex \emph{higher} than their values obtained from fitting the full UVOIR data. These galaxies correspond to the models in the lower left of the middle panels of Figure~\ref{fig:sps_results_compare}, where the sSFR is over-estimated in extremely low sSFR systems. These systems behave in the opposite fashion to their intermediate-sSFR counterparts, as their red colors are due to extremely old stars but the prior on dust reddening forces both their fitted sSFR and dust reddening values to be too high. UV data breaks this degeneracy and allows the extremely low sSFR and low dust extinction to be correctly recovered.

These trends observed in the photometry fitting of both the models and SDSS galaxies were also observed (albeit with smaller statistics) in the \snf\ \snia\ host galaxy sample. For the 394 hosts with \galex\ photometry, the mean sSFR uncertainty increased from 0.27~dex to 0.43~dex when the UV data was not included. We found the \galex\ UV data to be critical for correctly determining the sSFR of our massive elliptical hosts. For the 66 \snf\ host galaxies with $\log(M_*/M_\odot) > 10.0$ and $\log(sSFR) < -11.0$ (note we softened the cuts to increase the sample size), the sSFR values were on average 0.40~dex higher when \galex\ UV data was excluded in the photometry fit. Of those hosts, 35 have \galex\ data with lower than $5\sigma$ detections, illustrating that even modest detections or upper limits of flux in the UV provide critical information for the determination of \snia\ host galaxy star formation intensity.

\section{\snia\ Host Galaxies and the Galaxy Mass--Metallicity Relation}
\label{sec:snia_host_MZ}
The level of agreement of \snia\ host galaxies with the normal galaxy mass-metallicity (MZ) relation can provide important insight into preferred \snia\ progenitor environments. Discrepancies between the \snia\ MZ distribution and that of normal galaxies could potentially indicate metallicity preferences for \sneia, which would have important implications for high redshift \snia\ surveys. Alternatively, disagreement with the MZ relation could have other interpretations, as was the case with long-duration gamma ray burst host galaxies. 

Some recent studies of the host galaxies of long-duration gamma ray bursts (LGRBs) found that they tended to have systematically lower metallicities than those predicted by fiducial galaxy MZ relation \citep{modjaz08,levesque10b}. Initial interpretations of this trend speculated on a preference for lower metallicity environments in the production of LGRBs. The key insight, however, came from considering the effect of galaxy star-formation rate (SFR) on the galaxy MZ relation \citep{mannucci10a}. Accounting for this effect, it was found that LGRB hosts indeed agreed with the SFR-adjusted MZ relation (or equivalently the M-Z-SFR relation) but merely appeared in the region of galaxy parameter space populated by the most intensely star-forming galaxies \citep{kocevski11,mannucci11}. Thus this trend showed the preference for LGRBs to form in very young stellar environments.

The \snia\ host galaxy agreement with the MZ relation has been an implicit assumption of previous authors who interpreted \snia\ brightness trends with host galaxy stellar mass in terms of \snia\ progenitor metallicity. The \snf\ sample is ideal for testing this assumption, as our untargeted search found \sneia\ in an unbiased sample of host galaxies spanning a large range (nearly 5~dex) of stellar masses. In this Section we present our method for inspecting the consistency of \snia\ host galaxies with the galaxy MZ relation and the results from the hosts of \sneia\ discovered by \snf.

\subsection{\snf\ \snia\ Hosts and the MZ Relation}
The correlation of galaxy luminosity and stellar mass with metallicity has been known for several decades \citep{lequeux79}, but has been quantitatively refined only recently with the advent of major galaxy spectroscopic surveys at low \citep[SDSS][]{york00} and intermediate \citep{zahid11} redshifts. Of particular interest for this work is the correlation of galaxy stellar mass with gas-phase metallicity, which for simplicity we will refer to simply as ``metallicity'' in this Section. For SDSS the MZ relation was studied by the MPA-JHU SDSS team in \citet[][T04]{trem04} for the fourth SDSS data release and subsequently for future data releases. T04 found that for a sample of $\approx$45,000 galaxies, gas-phase metallicities followed a tight relation in the stellar mass range of $8.5 \leq \log(M_*/M_\odot) \leq 11.0$ with a dispersion of about 0.1~dex at high stellar masses. The dispersion in the MZ relation increases at lower stellar mass, up to about 0.3~dex at $\log(M_*/M_\odot) = 8.5$.

\begin{figure*}
\begin{center}
\includegraphics[width=0.90\textwidth]{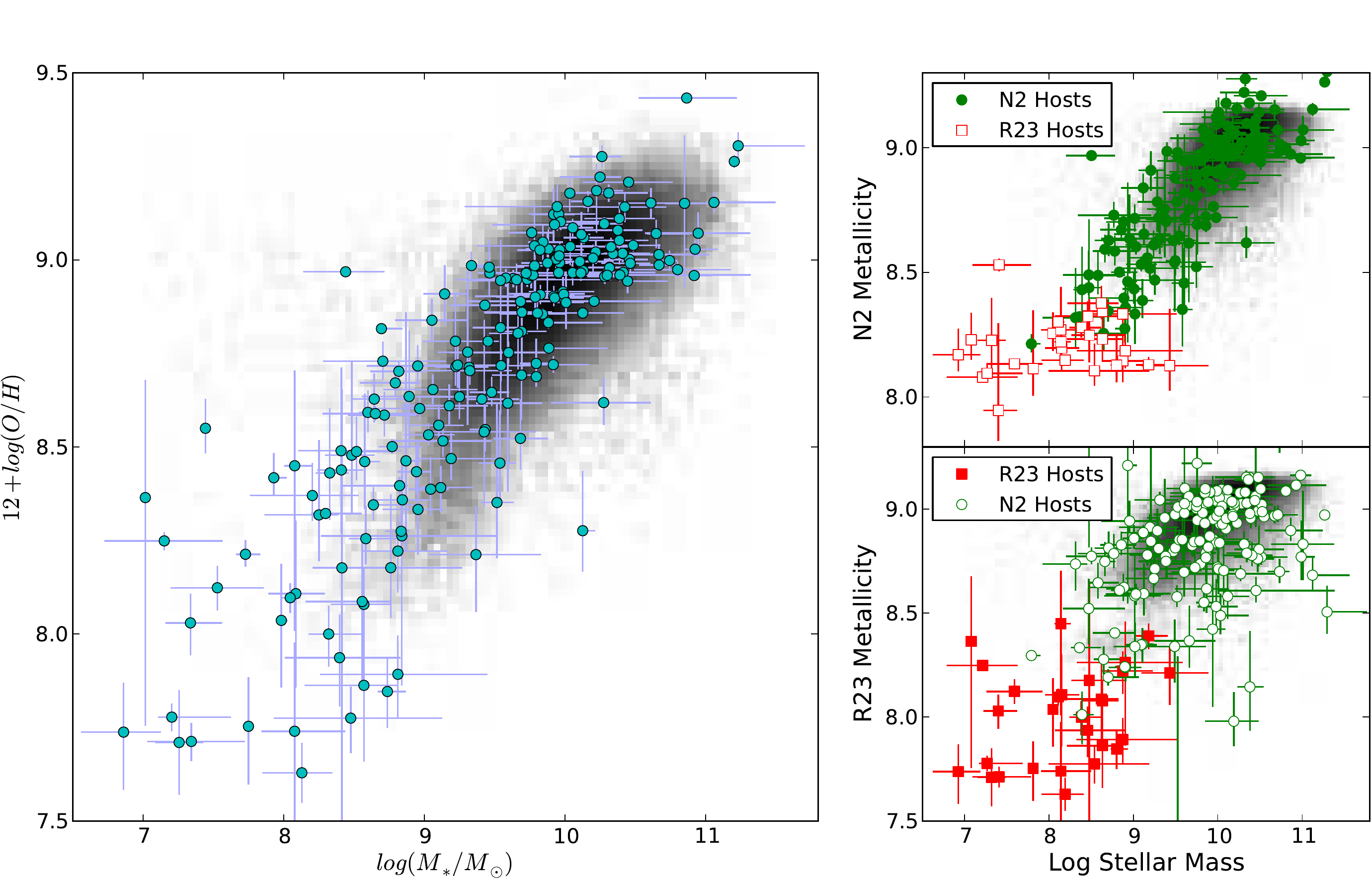}
\end{center}
\caption[\snf\ MZ Diagram]{Left: Location of \snf\ host galaxies in the MZ plane using final T04 metallicities. The grey background is a density plot of the galaxies in the SDSS DR7 sample analysis from the MPA-JHU team. Top Right: PP04 N2 metallicities for all \snf\ host galaxies. Those hosts whose final metallicities were combined using this method are shown as solid green circle, while those with final metallicities calculated from $R_{23}$ are shown as open red squares. Bottom Right: Same as top right, but showing $R_{23}$ metallicities for all \snf\ hosts, with filled red squares showing those hosts whose finally metallicities came from $R_{23}$ and open green squares those calculated from PP04 N2.}
\label{fig:full_snf_MZ}
\end{figure*}

For this analysis we wish to inspect how much \snia\ hosts deviate from the fiducial MZ relation and whether those deviations are consistent with the observed dispersion in the MZ relation. To do so we use derived stellar masses and metallicities from the MPA-JHU SDSS team analysis of the SDSS DR7 \citep{abaz09} data. They derive galaxy stellar masses from broadband photometry using the stellar population synthesis library of \citet{kauff03a}, and calculate gas phase metallicities from emission line fluxes according to the method outlined in T04. To facilitate the appropriate comparison, we use the \snf\ host stellar masses and metallicities derived in Sections~\ref{sec:snf_host_phot} and \ref{sec:snf_host_spec}. The MPA-JHU masses are computed with a \citet{kroupa01} IMF, which has a negligible mean offset from the \citet{chab03} IMF employed in our methods, though there exists some small scatter. As stated above, our metallicity values have all been converted to the T04 scale, so our masses and metallicities can be compared directly to those derived for SDSS galaxies by the MPA-JHU group. The full \snf\ host MZ diagram is shown in Figure~\ref{fig:full_snf_MZ}.

In order to assess the agreement of \snf\ host masses and metallicities with the SDSS MZ relation, we first compare the observed metallicity of each \snia\ host galaxy with the value predicted by the MZ relation for its observed mass. In practice, we calculate the weighted mean of the metallicities of all neighboring (in mass) SDSS galaxies with masses $M_i$, weighted by their distance from the observed host mass $M_\mathrm{host}$ (i.e. $\exp[-\chi_i^2/2]$ where $\chi_i^2 = ((M_i-M_\mathrm{host})/\sigma_M)^2$ and $\sigma_M$ is the uncertainty in the \snia\ host mass) with proper accounting for the number of SDSS hosts as a function of mass. Thus for each \snia\ host we can calculate the difference between its observed metallicity and that predicted from the MZ relation as $\Delta Z = Z_{host} - Z_{MZ}$, with an uncertainty equal to the quadrature sum of the host metallicity measurement error and the dispersion of the MZ relation at that host mass (i.e. the RMS of the metallicity values of its stellar mass neighbors from SDSS). 

Performing this calculation for all 130 \snia\ hosts in the \snf\ sample in the stellar mass range over which the MZ relation is well populated by SDSS, i.e. the aforementioned $8.5 \leq \log(M_*/M_\odot) \leq 11.0$, we find the weighted mean (and RMS) deviation of \snia\ host metallicities from the MZ relation to be:
\begin{equation}
  \left<\Delta Z\right> = 0.0007 \pm 0.1522 \nonumber
\end{equation}
For this number of hosts, the error on the mean is 0.012~dex. If we rephrase the MZ deviation in terms of pull values we find a mean and RMS deviation of:
\begin{equation}
  \left<\frac{\Delta Z}{\sigma_Z}\right> = 0.00 \pm 1.12 \nonumber
\end{equation}
Thus we can see that \snia\ host galaxy metallicities are, on average, very consistent with the values predicted for their host masses by the galaxy mass-metallicity relation. The fact that the RMS of our pull values is close to 1 implies agreement of \snia\ hosts with the MZ relation for not only the average metallicity values, but also the observed dispersion. This result also implies that our measurement error bars are not misestimated.

In the right panels of Figure~\ref{fig:full_snf_MZ} we plot the \snf\ host galaxy MZ relation for the $R_{23}$ and N2 metallicity metrics from which the final T04 metallicities were derived (see Section~\ref{sec:snf_host_metal}). These plots illustrate the aforementioned sensitivity failure of the NII line at low metallicities, as well as the double-valued nature of the $R_{23}$ metric. As can been seen from these plots, nearly all of the \snia\ host metallicities used in the MZ agreements tests of this Section come from the N2 metric, and show a similar agreement to the MZ relations of the SDSS sample using that metric.

\begin{figure*}
\begin{center}
\includegraphics[width=0.90\textwidth]{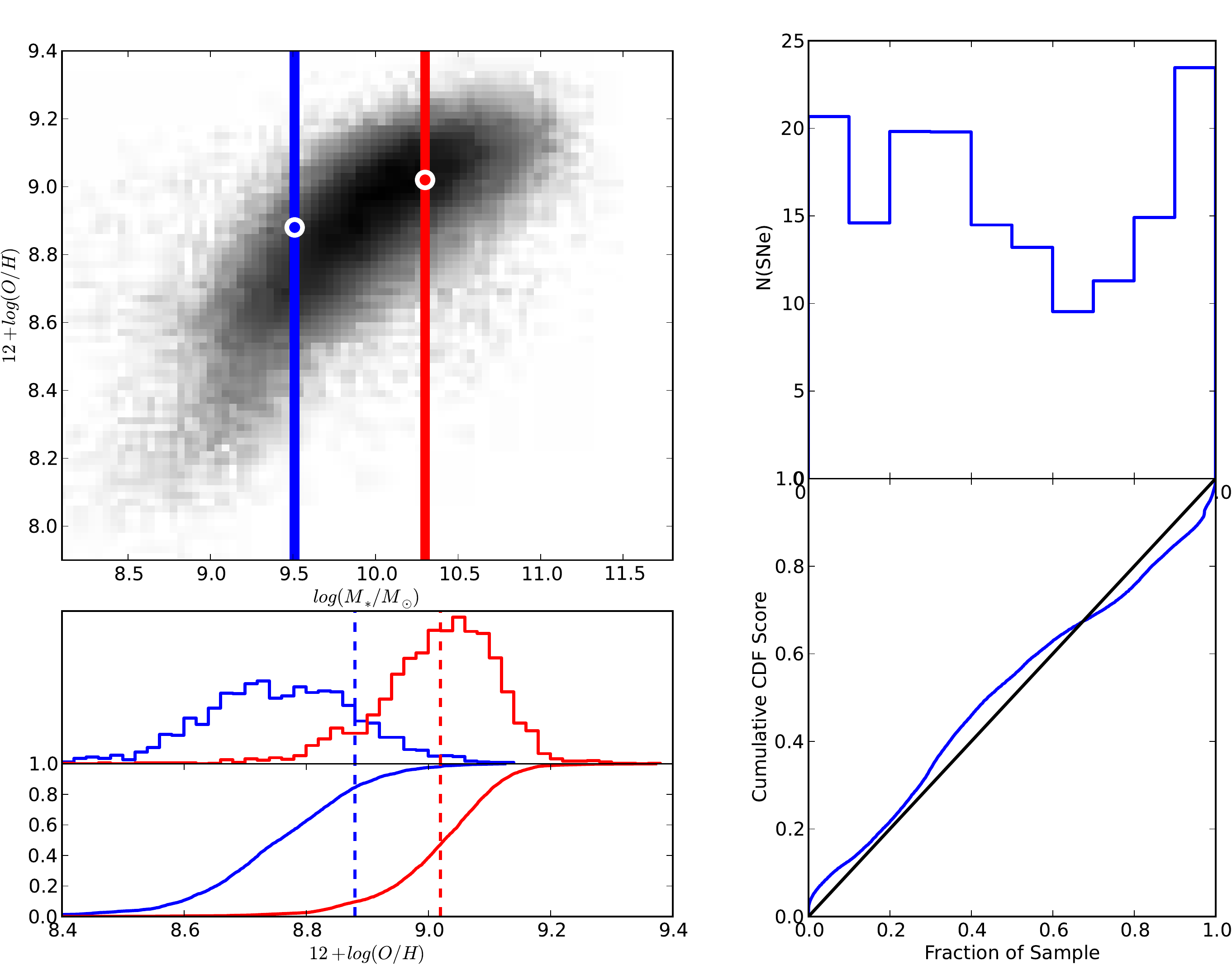}
\end{center}
\caption{Top Left: Example of method for calculating the MZ agreement score for each \snia\ host in the \snf\ sample. The blue and red bars correspond to the $\pm1\sigma$ mass values for two hosts, with the white circles showing their mass and metallicity values. Middle Left: The (unweighted) histogram of metallicities within $\pm0.05$~dex of each host mass. Bottom Left: The (weighted) cumulative distribution function (CDF) for metallicities at each host's mass. The score for each host is the intersection of its metallicity value (vertical lines) with the metallicity CDF at its mass. Top Right: Distribution of MZ scores for \snia\ hosts from \snf. Bottom Right: CDF comparison of the \snia\ host MZ score distribution to a uniform (flat) distribution.}
\label{fig:MZ_CDF_example}
\end{figure*}

To calculate each individual host galaxy's deviation from the SDSS MZ relation in further detail, we derive the metallicity cumulative distribution function (CDF) at each value of stellar mass, again using the weighted metallicities of each host's neighbors in stellar mass. The host is then assigned a score corresponding to where its metallicity is placed in the CDF of metallicities at its mass. This principle is illustrated in the left panels of Figure~\ref{fig:MZ_CDF_example}.

Thus for each \snia\ host galaxy, we have a measure of where its metallicity lies in the distribution of metallicities at its stellar mass, which we will call its MZ agreement score. If \snia\ hosts perfectly obey the MZ relation, then the ensemble distribution of these scores should be distributed uniformly between 0 and 1. We show in the right panels of Figure~\ref{fig:MZ_CDF_example} this distribution of MZ agreement scores for the 130 \snf\ \snia\ host galaxies whose mass falls within the aforementioned range. From this histogram we can see that the scores are relatively uniform. In the right panel of the same Figure, we plot the cumulative distribution function of the MZ agreement scores as compared to a line of unit slope (i.e. the CDF for a flat distribution). We can see from this plot that indeed our distribution is very close to a flat (uniform) distribution, and the cumulative distribution is reasonably close to unity. This would imply that not only are the mean and RMS metallicity deviation for \snia\ hosts consistent with the MZ relation, but the shape of their distribution is also similar.

We note that 30 hosts in our sample have host masses below the lower mass limit over which the MZ relation is well defined. We measure the RMS of the metallicity values of \snf\ hosts at the low mass end to be 0.32~dex, which is consistent with the 0.3~dex dispersion measured by T04 and the 0.4~dex dispersion measured by \citet{zahid12} at comparably low masses. Interestingly there appears to be a paucity of hosts below $12+log(O/H)=7.7$, near the low metallicity threshold predicted by \citet{kn09}. While interesting, we caution that proper interpretation of this result requires a complete accounting of both observational incompleteness and SN search efficiency. A careful treatment of this is underway and will be presented in a future \snf\ paper.

\section{Star-Formation Activity of \snia\ Host Galaxies}
\label{sec:snia_host_sfr}
In a similar spirit to the analysis of Section~\ref{sec:snia_host_MZ}, we wish to investigate whether the star-formation rates of \snia\ host galaxies are distributed similarly to the values one would measure from a comparable sample of normal field galaxies. Galaxy SFRs are not tightly correlated with stellar mass as was true for gas-phase metallicities, so here we will examine whether the \emph{distribution} of star-formation indicators for the \snf\ \snia\ host galaxy sample is similar to the same distribution for a comparable sample of field galaxies. 

This analysis begins with the expectation of finding subtle differences between \snia\ hosts and field galaxies driven by the \snia\ delay time distribution (DTD) -- the distribution of times from formation of the progenitor system to explosion as a \snia. \snia\ rates show a dependence on both host galaxy SFR and stellar mass \citep{mannucci05, scan05, mannucci06, sullivan06, aubourg08, smith12}, and studies of the \snia\ DTD indicate a larger fraction of young ($\lesssim 1$~Gyr) progenitors \citep{totani08, brandt10, kbar12, maoz12}. Therefore, we begin this comparison with the expectation that our \snia\ host galaxies may show enhanced star formation.

In order to compare \snia\ host galaxy SFRs to a sample of typical galaxies, we require an appropriate field galaxy sample -- ideally one with stellar masses and SFRs measured from photometry in a manner comparable to our own methods, as well as spectroscopic measurements of galaxy \ha\ emission. While there are many large well-studied samples that come close (e.g., NFGS \citep{jansen00}; 11HUGS \citep{kennicutt08, lee11}), the galaxy catalog that currently provides the closest match to our needs is again the MPA-JHU value-added SDSS catalog. We will use this as our baseline for establishing the SF properties of a field galaxy sample, but in what follows we will describe a number of systematic limitations that do not yet allow us to exploit the full statistical power of our \snia\ host galaxy sample.  Thus, the aim of this Section will be to test whether the star-formation activity of \snia\ host galaxies differs from that of a comparable sample of field galaxies at a level significantly above the systematic uncertainties.

Accordingly, we proceed by first discussing some key differences between the MPA-JHU method and our own technique for estimating SFRs from photometry in Section~\ref{sec:sfr_method_diffs}. Then in Section~\ref{sec:sfr_sample_weights} we discuss the importance of appropriately weighting the SDSS spectroscopic sample to account for SFR-dependent Malmquist bias. Finally in Section~\ref{sec:sfr_comparison_results} we present the results of our \snia\ host galaxy comparison to SDSS field galaxies.

\subsection{Differences in the SDSS and \snf\ SFR Measurement Methods}
\label{sec:sfr_method_diffs}
Photometric SFRs for the MPA-JHU SDSS catalog were presented in \citet[][hereafter B04]{brinchmann04}. These SFRs comprise two components: the first calculated directly from the spectrum obtained from the 3\arcsec\ fiber placed on the center of the galaxy, and the second calculated from the flux and colors of the galaxy light outside the fiber.

The fiber SFR values in B04 were calculated from a Bayesian technique similar to the photometric mass estimates applied to the measured emission line fluxes in the galaxy spectrum. These measured line fluxes were compared to a suite of emission line models from \citet{cl01} which spanned various values of metallicity, ionization parameter, and reddening by dust (note these are the same models and comparison methods used to calculate metallicities in T04). B04 found the final SFRs to be driven primarily by the flux in \ha, but with a scaling parameter that varied subtly from the \citet{kenn98} value depending on the best fit metallicity and ionization parameter.

SFR values for the galaxy flux outside the SDSS fiber were calculated in B04 using a color-color SFR grid trained on the spectroscopic data. Specifically, B04 calculated the SFR per unit luminosity in $i$-band within the SDSS fiber for all galaxies in their sample, then measured the mean $SFR/L_i$ in small bins in $g-r$ vs. $r-i$ color space. For each galaxy, the $g-r$ and $r-i$ colors of the light outside the SDSS fiber was measured and the mean $SFR/L_i$ for the corresponding color-color bin applied to the $i$-band flux outside the fiber.

Thus the B04 SFR values for the SDSS galaxy sample were trained to show tight agreement with SFRs measured from \ha\ flux (indeed they are partly composed of the \ha\ flux-based SFR from within the spectroscopic fiber). This explains the tight agreement between their photometric SFRs and SFRs calculated from \ha\ flux with the \citet{kenn98} relation (Figure~\ref{fig:ha_vs_phot_sfr}).

We now turn to the systematic differences between the B04 SFR technique and that employed by us (Section~\ref{sec:host_phot_sps}) for \snia\ host galaxies. Our technique measures the SFR of a galaxy from its average sSFR over the past 500~Myr, and is strongly driven by UV flux. This inherently probes a different timescale in the galaxy SFH than the \ha-driven SFRs of B04, which probe the very recent ($\lesssim$10~Myr) part of the galaxy SFH. There is cause to believe that these SFRs should agree on average but have some scatter due to the variety of SFHs realized in nature. This was borne out by the work of \citet{salim07} -- who performed a SED matching technique similar to, and the inspiration for, our own technique -- who found that the photometrically measured SFRs agreed with the \ha-driven SFRs with a scatter of 0.5~dex. Their SFRs were calculated over the last 100~Myr in their models, and the scatter is likely a product of the diversity of galaxy SFHs. Similar scatter (perhaps even larger) is to be expected when comparing our 500~Myr averaged SFRs to the \ha-driven ones of B04.

Differences between our SFRs and those derived with the B04 technique may also arise due to the fact that their values are derived (in part) from optical photometry. This could yield a sSFR-correlated bias as found in our models in Section~\ref{sec:filter_coverage}, and may explain the structure in the SFR comparison presented in Figure~\ref{fig:mpa_zpeg_compare}. These concerns, along with the SFH diversity scatter described above, limit the power of direct comparison between SFR values derived with these differing method. Our objective in this Section then will be to look for strong deviations in the SFR behavior of \snia\ hosts which cannot be explained by these systematic differences, while a detailed quantitative comparison will be best suited for a future analysis where consistent methods can be applied to \snia\ hosts and a large field galaxy sample.

\subsection{Weighting SDSS Galaxies for Selection Bias}
\label{sec:sfr_sample_weights}
The objective of this section is to the compare the distribution of \snia\ host galaxy SFRs, as traced by several indicators, to a comparable field galaxy sample from the MPA-JHU SDSS catalog. Unlike our previous analysis of the galaxy MZ relation, this SFR comparison must account for observational incompleteness resulting from the magnitude limit ($m_r=17.77$) of the SDSS spectroscopy survey. More intensely star-forming galaxies will be bluer and more luminous (i.e. have a lower mass-to-light ratio) than less intensely star-forming galaxies of the same mass. This means that the raw demographics of galaxies in the SDSS spectroscopic survey will be biased toward more strongly star-forming galaxies. This effect was not important in our consideration of the galaxy MZ relation (Section~\ref{sec:snia_host_MZ}) since metallicity does not strongly affect the galaxy luminosity, but it is of critical importance for the analysis of this Section.

To mitigate the effect of luminosity bias, we calculate for each galaxy in the MPA-JHU sample the maximum possible redshift $z_{lim}$ at which it would have been included in the SDSS spectroscopic sample, then calculate the effective volume of the SDSS sample in which these galaxies would be found ($V \propto (z_{lim}^3-z_{min}^3)$ where $z_{min}=0.02$ is the low-redshift cut for our sample -- see below). Each galaxy then is assigned a weight according to its effective fraction of the total survey volume ($V/V_{max}$ where $V_{max} \propto (z_{max}^3-z_{min}^3)$).

This $V/V_{max}$ weighting scheme is complicated by the fact that the SDSS spectroscopic sample includes some galaxies whose luminosity is below the nominal magnitude limit of the survey. These are typically low mass galaxies whose blue colors and small physical sizes made them prime candidates to be high-redshift quasars, which were spectroscopically targeted below the nominal magnitude cut. Indeed, below galaxy mass of $\log(M_*/M_\odot) \sim 8.5$ there are almost no galaxies which are more luminous than the nominal SDSS spectroscopic magnitude limit given our lower redshift cut. For completeness we wish to include these galaxies in order to characterize the SFR activity at low galaxy mass scales, but we must weight them differently than galaxies brighter than the nominal SDSS magnitude limit. Upon inspection of the magnitudes of these galaxies, we found that most of these galaxies obeyed an apparent upper magnitude limit in $r$-band of $m_r=20$. We therefore weighted all galaxies above the nominal SDSS spectroscopic magnitude limit of $m_r=17.77$ using a $V/V_{max}$ weight calculated using $m_r=20$ as the survey magnitude limit.

Since galaxy star-formation intensity (namely sSFR), correlates with galaxy mass \citep{salim07}, we construct the distributions of star-formation indicators from the SDSS sample using final weights that yield a weighted galaxy mass distribution equivalent to that of our \snia\ host galaxy sample. First, we employed redshift cuts of $z_{min}=0.02$, to mitigate the effects of low fiber filling fractions for massive galaxies at low redshift, and $z_{max}=0.10$, above which the SDSS spectroscopic survey is heavily biased towards massive luminous galaxies. $V/V_{max}$ volumetric weights were calculated for each galaxy using these values and the observed galaxy luminosity in $r$-band as described above. We then constructed the volume-weighted galaxy mass distribution for this sample and calculated a mass-based weighting function to force the final weighted mass distribution to match that of the \snia\ host sample.

In the analysis of Section~\ref{sec:sfr_comparison_results}, we will examine SF indicators for two subsets of the \snf\ sample: (i) those with SFRs (and sSFRs) measured from photometry, and (ii) those with \ha\ SFRs measured from spectroscopy. Since these are two distinct samples with different mass distributions, we calculated two mass-based weighting functions for the SDSS sample to make the appropriate comparisons to the respective samples.

\subsection{Comparison of \snia\ Host Galaxy SFRs with SDSS Field Galaxies}
\label{sec:sfr_comparison_results}
\begin{figure}
\begin{center}
\includegraphics[width=0.45\textwidth]{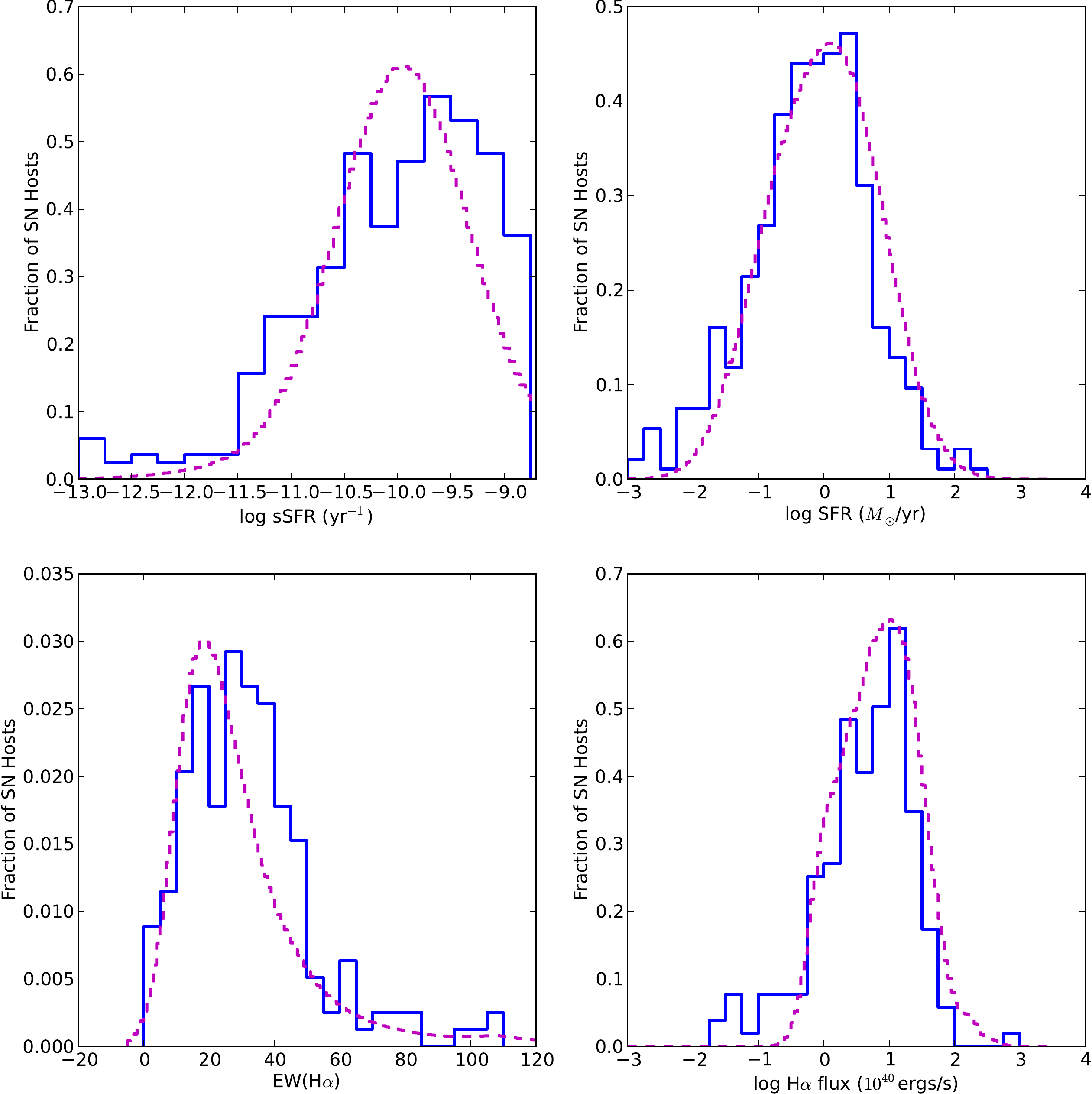}
\end{center}
\caption{Top Panels: Photometric estimates of sSFR (left) and SFR (right) for \snf\ \snia\ host galaxies (solid blue histograms) compared to the weighted distribution of those quantities from a comparable sample of SDSS galaxies (dashed magenta curves). Bottom: Spectroscopic measurements of the equivalent width of (left) and total flux in (right) the H$\alpha$ line, with curves the same as top panels.}
\label{fig:sfr_comparisons}
\end{figure}

Figure~\ref{fig:sfr_comparisons} presents a graphical comparison of star-formation indicators for the \snf\ \snia\ host galaxy sample compared to the appropriately weighted distribution of those same quantities from the MPA-JHU SDSS galaxy sample. The top left and right panels show sSFR and SFR measured from photometry, while the bottom left and right panels show the equivalent width of the \ha\ line and total galaxy flux in the \ha\ line. For this Figure the sSFR and SFR values from the MPA-JHU catalog have been randomly perturbed by values gaussianly distributed about 0 with a dispersion of 0.5~dex, in order to simulate the effect of the UV-\ha\ SFR dispersion observed by \citet{salim07} which we suggested above (Section~\ref{sec:sfr_method_diffs}) is likely to arise from SFH diversity. We note that the sSFRs for our \snia\ hosts are averaged over the last 500~Myr of the model SFHs and thus by construction have a maximum value of $\log(sSFR) = -8.7$ corresponding to the entirety of the galaxy mass being formed in this time interval.

In Table~\ref{tab:sfr_agreement} we summarize the properties of the star-formation indicator distributions for the SDSS galaxy sample and the \snf\ \snia\ host galaxies (we note that for this comparison we restrict the sample only to \sneia\ discovered by \snf). We present the median and $\pm1\sigma$ values for the two samples, but due to the systematic differences in photometric SFR measurement techniques (see Section~\ref{sec:sfr_method_diffs}) and the uncertain impact of aperture effects (see Section~\ref{sec:snf_host_spec}), we purposfully avoid power statistical tests such as a Kolmogorov-Smirnov test because they cannot effectively account for these subtleties.

\begin{table}
\begin{center}
\caption{Properties of SF Indicator Distributions}
\label{tab:sfr_agreement}
\begin{tabular}{lcrrr}
\hline
Quantity          & Sample & $-1\sigma$ & Median   & $+1\sigma$ \\
                  &        & Value      & Value    & Value      \\
\hline
log(sSFR) -- phot & SDSS   & $-10.61$   &  $-9.95$ & $-9.29$    \\
                  & SNf    & $-10.89$   &  $-9.93$ & $-9.18$    \\
log(SFR) -- phot  & SDSS   &  $-0.86$   &   $0.00$ &  $0.82$    \\ 
                  & SNf    &  $-1.12$   &  $-0.18$ &  $0.59$    \\
$EW(H\alpha)$     & SDSS   &   $13.6$   &   $24.2$ &  $47.0$    \\
                  & SNf    &   $13.9$   &   $30.0$ &  $46.9$    \\

$\log(F(H\alpha))$      & SDSS   &  $40.21$   &  $40.86$ & $41.42$    \\
                  & SNf    &  $39.92$   &  $40.72$ & $41.27$    \\

\hline
\end{tabular}
\end{center}
\end{table}

From this analysis we can make some general statements about the consistency of \snia\ host galaxy star-formation activity compared to a comparable sample of field galaxies. The differences between photometric SFR and sSFR distributions is much smaller than the systematic uncertainties in their measurements, so we detect no significantly unusual star-formation activity in \snia\ host galaxies as measured from photometry.

Interestingly, the $EW(H\alpha)$ distribution for \snia\ host galaxies appears to have a higher median value than the SDSS field galaxy sample. This could perhaps indicate that \sneia\ have a slight preference for more strongly star-forming environments. This would be consistent with recent studies of the \snia\ delay time distribution \citep[DTD -- e.g.][]{maoz11, kbar12} which find a DTD peaked at young stellar ages. Quantitative results from this analysis are limited by the uncertain impact of aperture effects, but $EW(H\alpha)$ provides the most interesting distinction between the \snia\ host galaxy and normal field galaxy samples.

In summary, \snia\ host galaxies show mostly normal star-formation activity given the current limitations on measurement of SFRs in our host galaxy sample and the field galaxy sample from SDSS. A more precise quantitative comparison will likely require a field galaxy sample whose properties are derived in a consistent way as the \snia\ hosts, and whose selection function is very well understood. Since \snia\ hosts have a higher relative fraction of low mass galaxies compared to the magnitude-limited SDSS sample, this comparison would also benefit from a field galaxy sample which is better sampled at the low mass end. For now we find modest agreement between the star-formation activity of \snia\ host galaxies and a comparable (in mass) sample of field galaxies.

\section{Summary}
\label{sec:summary}
Here we revisit the major contributions of this work. Specifically, we summarize the properties of the \snf\ \snia\ host galaxy sample compared to other \snia\ host samples, and illustrate that the \snf\ \snia\ sample found \sneia\ in environments missed by other surveys. We then discuss the advantages of our galaxy stellar population synthesis (SPS) technique, as well as its possible limitations. Finally we discuss the implications of our findings that \snia\ host galaxies exhibit similar metallicities and star-formation behavior as comparable samples of normal field galaxies.

\subsection{Advantages of the \snf\ \snia\ Sample}
The \snf\ \snia\ sample whose host galaxies are presented here represents a sample of 400 low redshift ($z \lesssim 0.1$) \sneia\ from an untargeted search, which we believe is the best available representation of the full population of \sneia\ and their host galaxies in the local universe. A representative statistic is the range of stellar masses probed by the sample of \sneia\ discovered by \snf, which spans from $\log(M_*/M_\odot)=6.9$ (SNF20080910-007) to $\log(M_*/M_\odot)=11.7$ (SNF20050926-002), with nearly half (46\%) of our host galaxies having stellar masses below $10^{10}M_\odot$. As a comparison, the compilation of local \snia\ host galaxies from \citet{neill09} spanned the mass range $8.6 \leq \log(M_*/M_\odot) \leq 12.24$, with only 15\% of their hosts having stellar masses below $10^{10}M_\odot$. 

We also used optical spectra to measure the gas phase metallicities for a large fraction of the \snf\ host galaxy sample, and showed that our hosts span nearly 2~dex in metallicity. This is especially advantageous for the construction of \snia\ spectral templates for application at high redshift where metallicities and stellar ages will be on average lower than in the low redshift universe. Indeed, a measurement of the galaxy mass-metallicity relation at $z\sim0.8$ by \citet{zahid11} showed that the MZ relation has shifted downward by approximately 0.15~dex, less than a tenth of the full metallicity range probed by our sample. Thus we can state that our ability to construct appropriate \snia\ templates at low redshift will not be hindered by cosmic chemical evolution since the range of metallicities probed locally far exceeds its average change at high redshift.

\subsection{\snia\ Host Galaxies and SPS Techniques}
In Section~\ref{sec:host_phot_sps} we presented our Bayesian technique for deriving \snia\ host galaxy masses and specific star formation rates from multi-band photometry. This method of applying stellar population synthesis (SPS) techniques to \snia\ host galaxies differs somewhat from those used by other authors studying \snia\ hosts \citep[e.g.][]{sullivan06, neill09, kelly10, sullivan10, lampeitl10}. We made several key observations about our technique which are important for future comparisons of \snia\ properties to the properties of their host galaxies.

First, we noted that SPS fits to galaxy photometry recovered the star-formation rate well only if UV data are available. Because of the degeneracy between redder colors in old stars and reddening by foreground dust, optical data alone cannot effectively distinguish between these two effects. Instead, both the galaxy sSFR and dust are underestimated at intermediate galaxy ages with modest reddening by dust, where the SPS recovery method mistakes the dust for older stellar ages. Thus our method is optimum when rest-frame UV photometry is available. For high-redshift ground-based surveys such as SNLS and DES, observer-frame optical photometry effectively samples the rest-frame UV, making these data unlikely to suffer biases in the estimate of host galaxy SFRs.

The second major result of our SPS analysis was that galaxy mass-to-light ratios (and thus final mass estimates) were recovered without significant bias even with limited photometric coverage. The reason for this is related to the aforementioned age--dust degeneracy. Both stellar age and foreground dust make stellar populations both redder and dimmer. While the ways in which the reddening and dimming effects are coupled (i.e. the slope of the respective color-luminosity relations) are not the same for age and dust, they are close enough that mass-to-light ratios are not made highly inaccurate. This serendipitous similarity between these effects means that the masses of \snia\ host galaxies can be fairly well recovered  from optical data alone. We also found that a single color results in significant uncertainty on the mass-to-light ratio of a galaxy, but this uncertainty is small compared to the range of masses probed by our \snia\ sample.

We then compared our mass and SFR estimates to those derived using other SPS techniques. A first comparison was applied to a sample of SDSS field galaxies whose masses and SFRs were measured by the MPA-JHU group using the methods of \citet{kauff03a}, which provided the intellectual template for our own methodology. We found excellent agreement with their values, as expected. We then used the code \zpeg, a popular code with \snia\ host studies, to derive masses for our hosts as well as the same SDSS training sample and a sample of our model galaxies. We found that the masses showed good agreement, but the fitted SFR values from \zpeg\ converged essentially to discrete values set by the input galaxy evolutionary scenarios. While such a discrete set of templates is the ideal set for deriving photometric redshifts, it provides coarse resolution for the detailed SFR measurements we desire.

Furthermore, such matching to small discrete SFH libraries provides an insufficient accounting for the uncertainty on a galaxy's SFH. While such limitations do not severely hinder the measurement of galaxy stellar masses, it provides a poor estimate of the \emph{uncertainty} on the mass. The inability of discrete SPS models to accurate quantify the uncertainty on galaxy physical properties was the prime motivation for our development of the SPS technique presented here. With the observation of a residual bias of \snia\ luminosities with host mass \citep{kelly10, sullivan10, lampeitl10, gupta11}, it is possible that host galaxy properties (especially stellar mass) may need to be included in \snia\ luminosity corrections in cosmological samples. In order to facilitate an appropriate accounting of systematic errors in this correction, we designed this SPS method to recover accurate measurement of the uncertainty on \snia\ host galaxy masses that accounts for the uncertainty in galaxy SFH.

Finally, we comment on some of the limitations of our SPS methods. Our estimates of SFR and mass recovery efficiencies were found when fitting our model spectra with other models from the same library, and thus can be considered a minimum level of systematic error associated with our technique. Our method is inherently tied a particular choice of SFHs which are meant to represent a reasonable prior on the likely distribution of galaxy SFHs in the local universe. Galaxies which are drawn from a different prior (i.e. if our prior is not an accurate representation of the true distribution of galaxy SFHs in the local universe) may have biases in the estimation of their physical properties \citep[see, e.g.,][]{gb09}. Similarly, our prior on the reddening of these galaxies and our choice of reddening law may introduce some biases if the true distribution of galaxy reddening or the reddening law differ from the choices made here. 

While a full assessment of the possible biases resulting from our choices is beyond the scope of this paper, we note the important fact that such biases are likely to affect all estimated galaxy properties in a coherent way. Thus a \emph{relative} comparison of galaxy properties amongst galaxies fitted with the same SPS techniques is entirely valid. It is for this reason that we chose the particular SPS techniques presented here, because they mimic very closely those employed by the major galaxy studies from the SDSS survey \citep[e.g.,][]{kauff03a, trem04, salim07}. Such self-consistency will also be important for \snia\ cosmological studies if \snia\ host galaxies are used for cosmological luminosity corrections.

\subsection{Comparison of \snia\ Hosts to Normal Galaxies}
In Sections~\ref{sec:snia_host_MZ} and \ref{sec:snia_host_sfr} we showed that the metallicity and star-formation activity of \snia\ host galaxies is very similar to that of a typical field galaxy sample from SDSS. Not only do the mean and RMS of \snia\ host metallicities agree with the normal galaxy MZ relation, but indeed the shape of the \snia\ host metallicity distribution is remarkably similar to that of normal galaxies. An  analysis of star formation indicators (total SFR and sSFR estimated from photometry, and H$\alpha$ equivalent width and total flux measured from spectroscopy) in \snia\ host galaxies showed that the distribution of those quantities in \snia\ hosts was similar to that calculated from a comparable SDSS field galaxy sample (though systematic differences in SFR measurement techniques between the two samples -- see Section~\ref{sec:sps_comparisons} -- limit the power of this comparison). This confirms that \sneia\ do not (on average) prefer extreme environments, but instead can be expected to be found in most normal galaxies.

The most important consequence of this result is that studies which investigate the variation of \snia\ properties with progenitor metallicity can be accomplished by means of measuring galaxy stellar masses from photometry, rather than the observationally expensive spectroscopic measurement of galaxy metallicity. Indeed several studies \citep{neill09, howell09} have previously attempted to measure the decrease in \snia\ \nifs\ yield predicted to occur at high metallicities \citep{tbt03} by measuring \nifs\ yields as a function of host galaxy mass. Similarly, the aforementioned studies which discovered the \snia\ luminosity host bias commented on the possibility of progenitor metallicity driving the effect. These and other works relied on the previous untested assumption that \snia\ host galaxies exhibit typical metallicity for their given stellar mass. We have shown here quantitatively that such assumptions were (retroactively) justified, and may indeed be employed in the future.

\section{Conclusions}
\label{sec:conclusions}
In this paper we presented photometric and spectroscopic observations of galaxies hosting \sneia\ discovered or observed by the Nearby Supernova Factory (\snf). Galaxy photometry in ten photometric filters spanning UV to NIR wavelengths were calculated for over 450 \snia\ host galaxies. Additionally, emission line fluxes from optical spectra were measured for over 300 \snia\ host galaxies. 

We presented a Bayesian method for fitting the \snia\ host galaxy photometry with stellar population synthesis (SPS) techniques inspired by many of the major SDSS galaxy sample studies. We illustrated the vital role \galex\ UV photometry plays in the correct estimation of galaxy star-formation intensity (sSFR) -- especially for galaxies with modest to low sSFRs -- in good agreement with the findings of previous studies such as \citet{neill09} and \citet{gupta11}. We further showed our methods to be consistent with other techniques for obtaining galaxy stellar masses, but with improved quantification of systematic uncertainties in SPS modeling.

Finally we used our galaxy physical parameters to compare the full sample of \snia\ host galaxies to a typical field galaxy sample from SDSS. We showed that \snia\ host galaxies obey the fiducial galaxy mass--metallicity (MZ) relation with remarkable agreement. We examined star-formation indicators in these \snia\ hosts and showed them to be distributed similarly as in an analogous sample of normal field galaxies. In summary, we have demonstrated that \snia\ host galaxies are extremely normal. 

\vskip11pt

\scriptsize
Acknowledgments: Based in part on observations made with the NASA Galaxy Evolution Explorer. \galex\ is operated fro NASA by the California Institute of Technology under NASA contract NAS5-98034. The authors graciously acknowledge support from \galex\ Archival Research Grant \#08-GALEX508-0008 for program GI5-047 (PI: Aldering).
This work was supported by the Director, Office of Science, Office of High Energy Physics, of the U.S. Department of Energy under Contract No. DE-AC02-05CH11231; the U.S. Department of Energy Scientific Discovery through Advanced Computing (SciDAC) program under Contract No. DE-FG02-06ER06-04; by a grant from the Gordon \& Betty Moore Foundation; in France by support from CNRS/IN2P3, CNRS/INSU, and PNC; in France by support from CNRS/IN2P3, CNRS/INSU, PNC, and Lyon Institute of Origins under grant ANR-10-LABX-66; and in Germany by the DFG through TRR33 ``The Dark Universe''. 
This research used resources of the National Energy Research Scientific Computing Center, which is supported by the Director, Office of Science, Office of Advanced Scientific Computing Research, of the U.S. Department of Energy under Contract No. DE-AC02-05CH11231.  We thank them for a generous allocation of storage and computing time. HPWREN is funded by National Science Foundation Grant Number ANI-0087344, and the University of California, San Diego. The Centre for All-sky Astrophysics is an Australian Research Council Centre of Excellence, funded by grant CE110001020.

The authors would like to thank the excellent technical and scientific staff at the many observatories where data was taken for this paper: the University of Hawaii 2.2m telescope, Lick Observatory, Keck Observatory, the Blanco 4m telescope, the SOAR telescope, and Gemini South. Some data presented herein were obtained at the W. M. Keck Observatory, which is operated as a scientific partnership among the California Institute of Technology, the University of California, and the National Aeronautics and Space Administration; the Observatory was made possible by the generous financial support of the W. M. Keck Foundation. We wish to recognize and acknowledge the very significant cultural role and reverence that the summit of Mauna Kea has always had within the indigenous Hawaiian community, and we are extremely grateful for the opportunity to conduct observations from this mountain. 
We also thank Dan Birchall for assistance with SNIFS observations.

Some of the data analyzed here were obtained from the Sloan Digital Sky Survey Eight Data Release (SDSS-III DR8). Funding for SDSS-III has been provided by the Alfred P. Sloan Foundation, the Participating Institutions, the National Science Foundation, and the U.S. Department of Energy Office of Science. The SDSS-III web site is http://www.sdss3.org/. SDSS-III is managed by the Astrophysical Research Consortium for the Participating Institutions of the SDSS-III Collaboration including the University of Arizona, the Brazilian Participation Group, Brookhaven National Laboratory, University of Cambridge, Carnegie Mellon University, University of Florida, the French Participation Group, the German Participation Group, Harvard University, the Instituto de Astrofisica de Canarias, the Michigan State/Notre Dame/JINA Participation Group, Johns Hopkins University, Lawrence Berkeley National Laboratory, Max Planck Institute for Astrophysics, New Mexico State University, New York University, Ohio State University, Pennsylvania State University, University of Portsmouth, Princeton University, the Spanish Participation Group, University of Tokyo, University of Utah, Vanderbilt University, University of Virginia, University of Washington, and Yale University.

Additional derived quantities for SDSS galaxies were obtained from the MPA-JHU database at http://www.mpa-garching.mpg.de/SDSS/ as derived from the Seventh Data Release (DR7) of SDSS. Funding for the SDSS and SDSS-II has been provided by the Alfred P. Sloan Foundation, the Participating Institutions, the National Science Foundation, the U.S. Department of Energy, the National Aeronautics and Space Administration, the Japanese Monbukagakusho, the Max Planck Society, and the Higher Education Funding Council for England. The SDSS Web Site is http://www.sdss.org/. The SDSS is managed by the Astrophysical Research Consortium for the Participating Institutions. The Participating Institutions are the American Museum of Natural History, Astrophysical Institute Potsdam, University of Basel, University of Cambridge, Case Western Reserve University, University of Chicago, Drexel University, Fermilab, the Institute for Advanced Study, the Japan Participation Group, Johns Hopkins University, the Joint Institute for Nuclear Astrophysics, the Kavli Institute for Particle Astrophysics and Cosmology, the Korean Scientist Group, the Chinese Academy of Sciences (LAMOST), Los Alamos National Laboratory, the Max-Planck-Institute for Astronomy (MPIA), the Max-Planck-Institute for Astrophysics (MPA), New Mexico State University, Ohio State University, University of Pittsburgh, University of Portsmouth, Princeton University, the United States Naval Observatory, and the University of Washington.

\bibliographystyle{apj}
\bibliography{snf_host_data}

\end{document}